\documentclass[a4paper,fleqn]{cas-dc}

\usepackage[numbers,sort&compress]{natbib}
\usepackage[utf8]{inputenc} %
\usepackage[T1]{fontenc} %

\def\tsc#1{\csdef{#1}{\textsc{\lowercase{#1}}\xspace}}
\tsc{WGM}
\tsc{QE}

\usepackage[separate-uncertainty=true,per-mode=symbol]{siunitx} %
\usepackage{nicefrac}
\usepackage{pifont} %
\usepackage{booktabs}
\usepackage{xcolor}

\newcommand{\alert}[1][]{}
\newcommand{\ryz}[1]{}
\newcommand{\rev}[1]{{#1}}

\newcommand{\equref}[2][]{Eq.\,\ref{#2}#1}

\newcommand{\secref}[2][]{Sec.\,\ref{#2}#1}
\newcommand{\figref}[2][]{Fig.\,\ref{#2}#1}
\newcommand{\Figref}[2][]{Figure\,\ref{#2}#1}
\newcommand{\mytabref}[2][]{Tab.\,\ref{#2}#1}

\begin{document}
\let\WriteBookmarks\relax
\def\floatpagepagefraction{1}
\def\textpagefraction{.001}

\shorttitle{FreDomLUS for fusion target characterization}    

\shortauthors{M.\ Ryzy et~al.}  

\title[mode = title]{Frequency domain laser ultrasound for inertial confinement fusion target wall thickness measurements}  

\author[1]{Martin Ryzy}[type=editor, %
	orcid=0000-0002-3632-2891]
\credit{Conceptualization (lead); Writing - original draft (lead); Writing – Review \& Editing (equal); Project administration (lead); Supervision (equal); Funding acquisition (equal); Investigation (equal); Software (lead); Visualization (lead); Methodology (lead)} %

\author[1]{Guqi Yan}[type=editor, %
	orcid=0000-0002-5029-2609]
\cormark[1] %
\ead{guqi.yan@recendt.at} %
\credit{Writing – Review \& Editing (lead); Investigation (lead); Data curation (lead); Software (equal); Visualization (equal)} %

\author[1]{Clemens Grünsteidl}[%
    type=editor, 
    orcid=0000-0001-5101-8479
]
\ead{clemens.gruensteidl@recendt.at} %
\credit{Writing – Review \& Editing (supporting); Validation (supporting)} %

\author[1]{Georg Watzl}[type=editor, %
    orcid=0000-0001-9883-0173
]
\ead{georg.watzl@recendt.at} %
\credit{Writing – Review \& Editing (equal); Validation (supporting)} 

\author[2]{Kevin Sequoia}[type=editor, %
]
\ead{kevin.squoia@ga.com} %
\credit{Writing – Review \& Editing (supporting); Resources (equal); Validation (supporting)} %

\author[2]{Pavel Lapa}[type=editor, %
]
\ead{pavel.lapa@ga.com} %
\credit{Writing – Review \& Editing (supporting); Resources (equal); Validation (supporting)} 

\author[2]{Haibo Huang}[type=editor, %
	]
\credit{Conceptualization (equal), Writing – Review \& Editing (supporting); Project administration (equal), Supervision (lead); Funding acquisition (lead); Resources (equal), Validation (supporting)} %

\affiliation[1]{organization={Research Center for Non Destructive Testing},
            addressline={Science Park 2 / 2.OG, Altenberger Straße 69}, 
            city={Linz},
            postcode={4040}, 
            state={},
            country={Austria}}
        
\affiliation[2]{organization={General Atomics, Energy Group, Inertial Fusion Technology Division},
        	addressline={}, 
        	city={San Diego},
        	postcode={92121}, 
        	state={CA},
        	country={USA}}
        
\cortext[1]{Corresponding author}

\begin{abstract} %
    In inertial confinement fusion experiments hollow, spherical mm-sized capsules are used as a container for nuclear fuel. 
    To achieve maximum implosion efficiency, a perfect capsule geometry is required. 
    This paper presents a wall thickness measurement method based on zero-group velocity guided elastic wave resonances. 
    They are measured with a non-destructive, contactless frequency domain laser ultrasound microscopy system. 
    Wall thickness measurements along the equator of a high-density carbon capsule with a diameter of around $\qty{2}{mm}$ and a wall thickness of around $\qty{80}{\micro\metre}$ excellently agree with infrared interferometry reference measurements. 
    In addition, the multi-resonant nature of a spherical shell is studied by complementing experimental observations with plate dispersion calculations and finite element wave propagation simulations. 
    The presented method is scalable and can be applied to a broad range of target materials, including metals, or metal-doped targets.
\end{abstract}

\begin{highlights} 
\item Resolved sub-micron thickness variations of mm-scale high-density carbon shells
\item Spherical shell acoustic spectrum exhibits plate- and circumferential-resonances %
\item Dispersion relation contains Lamb-modes superimposed by circumferential resonances %
\item Zero-group-velocity plate resonances can be isolated for wall thickness determination %
\item Finite element wave propagation simulation in spherical shell %
\end{highlights}


\maketitle

\section{Introduction}\label{sec:intro}
The implementation of nuclear fusion reactions under controlled laboratory conditions is one of the most challenging scientific and technological endeavors of our time. 
Yet it has the potential to significantly contribute to liberating us from the dependency on fossil fuel-based energy production. 
One of the most promising approaches in nuclear fusion research is inertial confinement fusion (ICF) \cite{hurricane_physics_2023}. 
Using that technique, a net target energy gain has been recently achieved in ignition experiments conducted at the National Ignition Facility (NIF) at the Lawrence Livermore National Laboratory in California, USA \cite{abu-shawareb_achievement_2024}.

The deuterium-tritium fuel forms a uniform ice-layer on the interior of a mm-sized spherical ablator target capsule made of high-density carbon (HDC) or other materials \cite{biener_diamond_2009, ross_high-density_2015, cardenas_progress_2018, braun_tungsten_2023}. 
To reach the extreme temperature and pressure to start a fusion reaction, the capsule is compressed by repulsive forces created by abrupt target material ablation initiated by X-ray irradiation. 
As this mechanism sets high demands on target material quality and symmetry \cite{braun_tungsten_2023,kritcher_design_2024,tollefson_how_2024}, precise, non-destructive characterization methods are required, which enable the selection of the most promising capsules for the high-yield ignition fusion experiments. 
One of the key parameters is the wall thickness uniformity of the capsules. 
HDC capsules have a thickness variation rejection criterion of \qty{0.35}{\micro\metre} out of an average wall thickness of \qty{80}{\micro\metre}. 
For optically transparent or translucent target materials, such as poly($\alpha$-methylstyrene), polystyrene and glow discharge polymer plastic capsules, the wall thickness can be determined by optical methods like infrared interferometry for instance \cite{caseyEvidenceThreeDimensionalAsymmetries2021}. 
However, for any opaque target materials like metals, or metal-doped materials, optical methods are limited due to the low optical penetration depth. 
As an alternative, X-ray imaging can be used for this class of samples. 
Due to limited sensor array size and the need to fit the entire \qty{2}{mm} capsule within the field of view, the ultimate wall thickness variation measurement precision is around $\pm\qty{0.15}{\micro\metre}$, or more than one third of the wall thickness nonuniformity allowed for HDC capsules. 
Considering that future shells can be bigger and made of higher-Z metals, not every capsule is imageable by X-ray transmission imaging. 
Thus, to pave the way for a broader range of target materials, alternative wall thickness measurement techniques are required.

Ultrasonic elastic waves also penetrate optically absorbing or reflective solids, and thus are suitable for gauging their thickness \cite{ scrubyLaserUltrasonicsTechniques1990,cheeke_nondestructive_2012}. 
A precise method for this is the measurement of resonance frequencies of guided elastic waves in plates or plate-like structures. 
They sustain an infinite number of laterally propagating wave-modes, which are characterized by strongly non-linear dispersion curves $f\!=\!f(k)$. 
Here, $f$ is the acoustic frequency and $k$ the angular (in-plane) wavenumber. 
In elastically isotropic materials such guided waves are denoted as Lamb-waves \cite{viktorov_rayleigh_1967, graff1991_p431}. 
Resonances occur at frequencies associated with local minima of the dispersion curves, i.\,e., at points where their group velocity $v_\mathrm{g}\!=\!2\pi\,\nicefrac{\partial{}f}{\partial{}k}$ vanishes. 
Depending on the elastic properties, in some dispersion branches, such minima can occur at points with a non-zero wavenumber (finite wavelength), and they are called zero-group velocity (ZGV) resonances \cite{tolstoy_wave_1957,gibson_lamb_2005,pradaLaserbasedUltrasonicGeneration2005,prada_local_2008}. 
Excited with a local acoustic source, their wave-motion remains confined in the excitation region, and they can thus be considered as local \cite{prada_local_2008}. 
As their frequency $f_\mathrm{ZGV}$ inversely scales with the local plate thickness $h$ \cite{graff1991_p431, every_intersections_2016}, relative thickness changes can be locally probed by measuring the ZGV resonance frequency $f_\mathrm{ZGV}$ and using the relation \cite{pradaLaserbasedUltrasonicGeneration2005}
\begin{equation}
	\frac{\Delta{}f_\mathrm{ZGV}}{f_\mathrm{ZGV}} = -\frac{\Delta{h}}{h}\mathrm{.}\label{eq:ZGV_scaling}
\end{equation}
Laser ultrasonic (LUS) techniques \cite{scrubyLaserUltrasonicsTechniques1990, spytek_non-contact_2023} are particularly well suited for that task, as they can provide local acoustic generation and detection \cite{pradaLaserbasedUltrasonicGeneration2005}. 
Based on that technique, thickness measurements of mm- \cite{watzl_simultaneous_2025} or sub-mm thick metalic plates \cite{Gruensteidl2018_v112_p251905}, imaging of \unit{\micro\metre}-range thickness variations in a silicon wafer \cite{balogun_high-spatial-resolution_2011}, or sub-micron layer on a substrate thickness variation measurements \cite{ces_thin_2011} have been demonstrated. 
Apart from that, a multitude of other laser ultrasonic, ZGV-based material characterization methods have been developed in the last two decades \cite{prada_power_2008,yan_estimation_2021,ryzy_determining_2023, thelen_laser-excited_2021,Yan2020_v116_p102323}, including elastic constant characterization \cite{clorennec_local_2007, ces_characterization_2012,grunsteidl_inverse_2016,watzl_situ_2022}. 
\rev{In recent years, the ZGV resonance has been applied for acoustoelastic characterization \cite{Morales2024} and damage assessment, such as delamination in composites \cite{Zhang2025}, local degradation in bonded plates \cite{Chen2026} and fatigue induced crack \cite{Lu2025}.}

Guided waves including ZGV-resonances have also been studied in curved plates like hollow cylinders, or spherically curved shells. 
If the relative thickness, which is the ratio of plate-thickness $h$ and outer radius $R$, is low, guided waves approximately behave like in straight plates. 
Neglecting circumferential modes, they can be described by Lamb's theory. 
The larger the relative thickness, the stronger the deviation is, and the effect is more pronounced for low frequencies (large wavelengths) \cite{lamb_vibrations_1882, towfighi_elastic_2003,clorennec_local_2007,ces_characterization_2012}. 
For instance, \`{C}es et al.\ showed that even for a hollow Zirconium alloy cylinder with a relative thickness of \qty{12}{\percent}, the first ZGV-resonance frequency deviates less than \qty{0.1}{\percent} from its straight-plate counterpart \cite{ces_characterization_2012}.

Most LUS techniques rely on pulsed laser excitation, with typical pulse durations ranging from the ns- to fs-scale. 
This results in broadband excitation of acoustic frequencies typically in the MHz to low GHz range, with some specialized system reaching into the low THz range \cite{scruby_5_1990,grahn_picosecond_1989,matsuda_fundamentals_2015}. 
However, generating acoustic signals with sufficient SNR ratio often requires sample ablation. 
But in cases where ablation damage is unacceptable, and only a narrow frequency band is required, e.\,g.\ when measuring single resonances, time-domain LUS may not be optimal. 
Frequency domain laser ultrasound (FreDomLUS) offers an alternative, which circumvents the mentioned issues by using a time-harmonic intensity modulated low power excitation laser in combination with phase sensitive detection. 
This way, narrow bands around an acoustic resonance can be probed directly in the frequency domain with a fine frequency resolution \cite{murray_high-sensitivity_2004, ryzy_measurement_2018, grunsteidl_evaluation_2020-1,  grunsteidl_measurement_2020, ryzy_determining_2023}. 
This technique was used to measure plate-resonances at GHz frequencies in fragile, micron-scaled plates for instance \cite{ grunsteidl_evaluation_2020-1,  grunsteidl_measurement_2020, ryzy_determining_2023}.

In this paper, we demonstrate ZGV-resonance based local wall thickness measurements in a spherical, hollow \qty{80}{\micro\metre} thin \ryz{W-doped} HDC-capsule with a frequency domain laser ultrasound (FreDomLUS) microscopy setup. 
In addition, broadband FreDomLUS measurements reveal the intriguing resonance landscape of the shell, that is analyzed with supporting finite element method (FEM) wave-propagation simulations, and numerical plate-dispersion calculations. 
We thereby identify local ZGV- and non-local circumferential resonances of the fundamental mode (SAW) and propose a simple method to isolate ZGV-resonances for local wall thickness determination. 
To the best of our knowledge, this is the first time that ZGV-resonance have been studied experimentally in a spherical shell.

\section{Materials and methods}\label{sec:methods}
\subsection{High-density carbon capsule sample}
The investigated sample is a high-density carbon (HDC) spherical shell with a nominal diameter of \qty{2}{mm} and a nominal wall thickness of $h\!=\!\qty{80}{\micro\metre}$ (see photograph and sketch in \figref[(a)]{fig:setup}). 
It contains a \qty{0.38}{at\%} W-doped internal layer with a thickness of about \qty{6}{\micro\metre}. 
The HDC originates from Diamond Materials GmbH (Freiburg, Germany) and was further processed by General Atomics (San Diego, USA) to yield the shell-geometry \cite{biener_diamond_2009}. 
The capsule diameter was measured with a micrometer screw gauge at ten random positions, yielding a diameter of $2R\!=\!\qty{2.254(6)}{\milli\meter}$, giving a relative thickness $\nicefrac{h}{R}\!\approx\!\qty{7}{\percent}$. 
For laser-ultrasonic experiments, the capsule is placed on a mechanical rotation stage, which enables us to perform ultrasonic measurements along the circumference of the shell. 
For the sake of simplicity, we neglect the W-doped layer in any numerical calculations (sections \ref{sec:disp} and \ref{sec:FEM}) \rev{
    --- its influence on ZGV frequency is estimated to be about \qty{0.05}{\percent}, which is much lower than the thickness variation of about \qty{1}{percent}, and we expect no significant change in scaling law (see Fig. S1 in supplementary in Sec. \ref{sec:suppl}) ---
} and use the elastic properties of pure HDC instead (see \mytabref{tab:HDC_props}).
\ryz{GA: details of sample fabrication}
\begin{table}
	\renewcommand{\arraystretch}{1.15}
	\centering
	\caption{Elastic properties of high-density carbon \cite{noauthor_diamond_2025, Klein1993diamond, Cho2005diamond} and calculated sound speeds\rev{, as used for the corresponding Lamb wave computation shown in \figref[a]{fig:broadband}}.} \label{tab:HDC_props}
	\begin{tabular}{  r c c}
		\toprule
		Youngs's modulus & $E$		& \qty{1050}{GPa} \\
		Poisson's ratio  & $\nu$	& \num{0.10} \\
		density			 & $\rho$	& \qty{3515}{kg\per\metre\cubed} \\
		\midrule
		longitudinal speed of sound 	& $c_\mathrm{L}$ & \qty{17479}{\metre\per\second} \\
		transverse speed of sound 		& $c_\mathrm{T}$ & \qty{11653}{\metre\per\second} \\
		Rayleigh speed of sound 		& $c_\mathrm{R}$ & \qty{10407}{\metre\per\second} \\
		\bottomrule
	\end{tabular}
\end{table}

\subsection{Frequency domain laser ultrasound experiments}
\begin{figure}
	\centering
	\includegraphics[width=\linewidth]{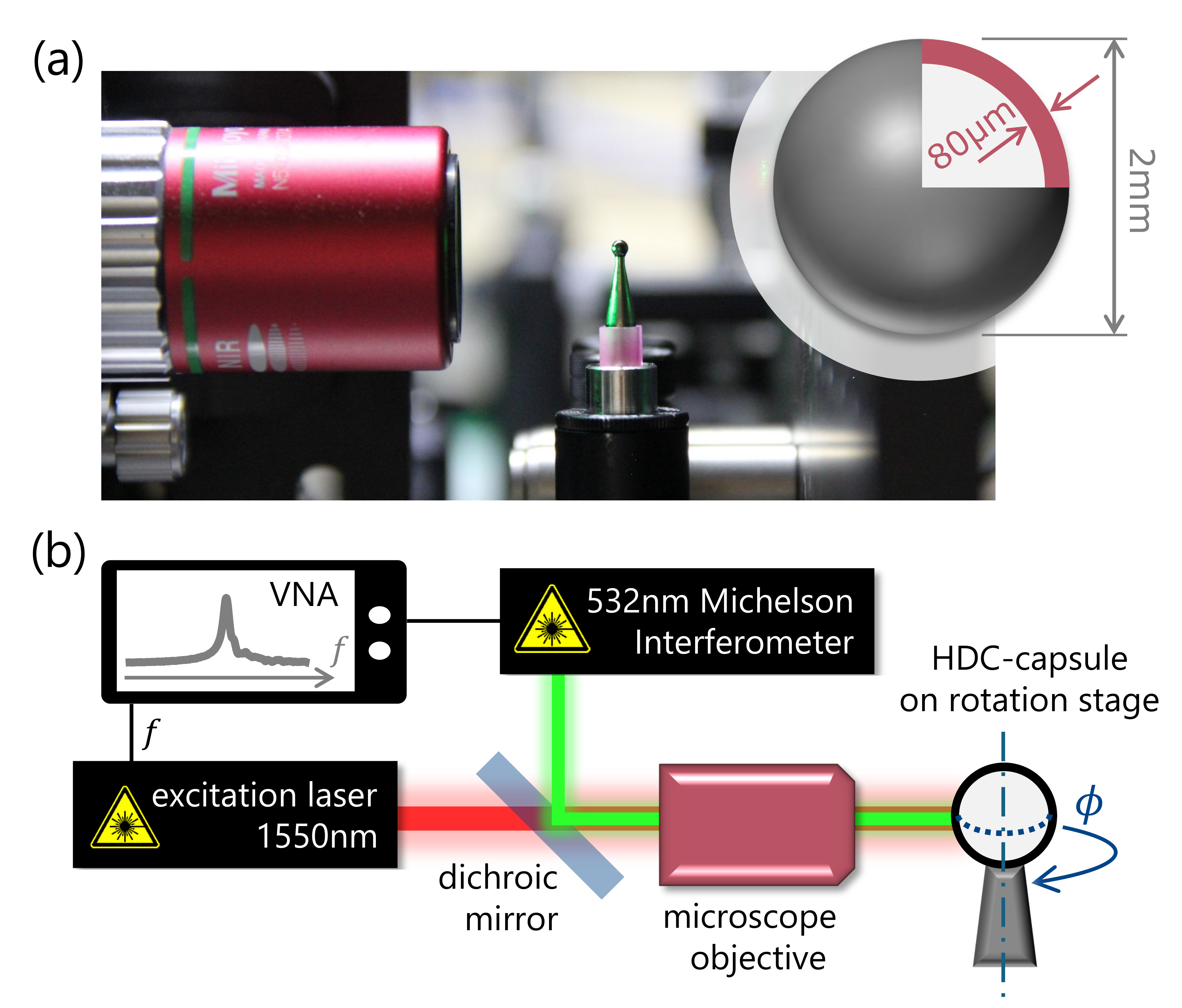}
	\caption{
        (a) Close-up of the HDC-capsule in front of the microscope objective and a sketch showing its geometry and dimensions (inset). 
        (b) Sketch of the FreDomLUS-Setup used for thermoelastic excitation and detection of the ultrasonic frequency response of an HDC-capsule. 
        The capsule is placed on a manual rotation stage for point-wise measurements along its circumference.\label{fig:setup}
    }%
\end{figure}
For the measurement of local acoustic response spectra, we use a frequency domain laser ultrasound microscopy setup (\figref[b]{fig:setup}), which has been described in more detail elsewhere \cite{murray_high-sensitivity_2004, ryzy_measurement_2018, grunsteidl_measurement_2020}. 
For thermoelastic excitation of ultrasonic waves, an electro-absorption intensity modulated laser diode with a wavelength of \qty{1550}{nm} is amplified by an erbium-doped fiber amplifier (EDFA) to about \qty{200}{mW} peak-to-peak power. 
It is focused by \rev{a} microscope objective to a spot with \rev{a full width at half maximum (FWHM) diameter of approximately \qty{2}{\micro\metre} \cite{grunsteidl_measurement_2020}} onto the HDC-capsule's surface. 
\rev{The thermal diffusion times around the laser spot radius can be estimated to be $\tau\!\approx\!(\qty{2}{\micro\meter}/2)^2 / \alpha_\mathrm{C}\!\approx\!\qty{2}{\nano\second}$, where $\alpha_\mathrm{C}\!\approx\!\qty{500}{\milli\meter\squared\per\second}$ \cite{Relyea1998} is the thermal diffusivity of diamond.
Therefore, purely thermoelastic excitation is expected for modulation frequencies not exceeding $1/\qty{2}{\nano\second}\!=\!\qty{500}{\mega\hertz}$ --- assuming no extreme buildup of heat in the bulk material.
}

A time-harmonic modulation signal is provided by a vector network analyzer (VNA). 
The laser source thermo-elastically couples into elastic waves sustained in the sample at the same frequency $f$. 
The induced surface normal displacement is detected by a path-stabilized Michelson-Interferometer \cite{scrubyLaserUltrasonicsTechniques1990_p76} which uses a \qty{532}{\nano\meter} continuous-wave laser that is co-aligned with the excitation laser. 
The initial \qty{120}{\milli\watt} detection laser is attenuated to \qty{13.5}{\milli\watt} with neutral density filters, and approximately half of the optical power impinges on the sample. 
The VNA records the Michelson's output in a phase-sensitive manner, yielding a complex voltage $\tilde{U}_\mathrm{VNA}(f)$ which is proportional to the surface normal displacement $\tilde{u}_r(f)$, where the subindex $r$ denotes the radial direction. 
To retrieve an acoustic response spectrum, the modulation frequency $f$ is \rev{swept, with step sizes of 50 kHz and 10 kHz for broadband and narrowband measurements, respectively.}. 
\rev{For precise extraction of ZGV frequencies, the measured frequency-domain spectra were first converted to the time domain using an inverse fast Fourier transform (iFFT).
A time gate was then applied to retain only early-arriving contributions, which suppresses circumferential resonances while preserving ZGV resonances. 
The gated signal was subsequently transformed back to the frequency domain using fast Fourier transform (FFT) to obtain the processed spectrum. 
Details of this time-domain gating procedure are provided in Sec. \ref{sec:resonances}.}

\subsection{Plate dispersion calculations}\label{sec:disp}
For the calculation of the elastic guided wave dispersion curves in straight HDC plates with the same thickness as the HDC-capsule, we use the M\textsc{ATLAB}-toolbox \texttt{GEWtool} developed by Kiefer et al.\ \cite{kiefer_gewtool_2023}. 
It is based on a semi-analytical spectral elements method (SEM) and enables a fast calculation of dispersion curves, i.\,e. of all pairs of acoustic frequency $f$ and in-plate-direction wavenumber $k$, which are sustained by the plate. 
In addition, the location of ZGV-points and the group velocity ($v_\mathrm{g}$) can be computed \cite{kiefer_computing_2023, kiefer_electroelastic_2025}. 
The isotropic elastic properties as stated in \mytabref{tab:HDC_props} have been used for all computations.

\subsection{Finite element modeling}\label{sec:FEM}
\begin{figure}
	\centering
	\includegraphics[width=\linewidth]{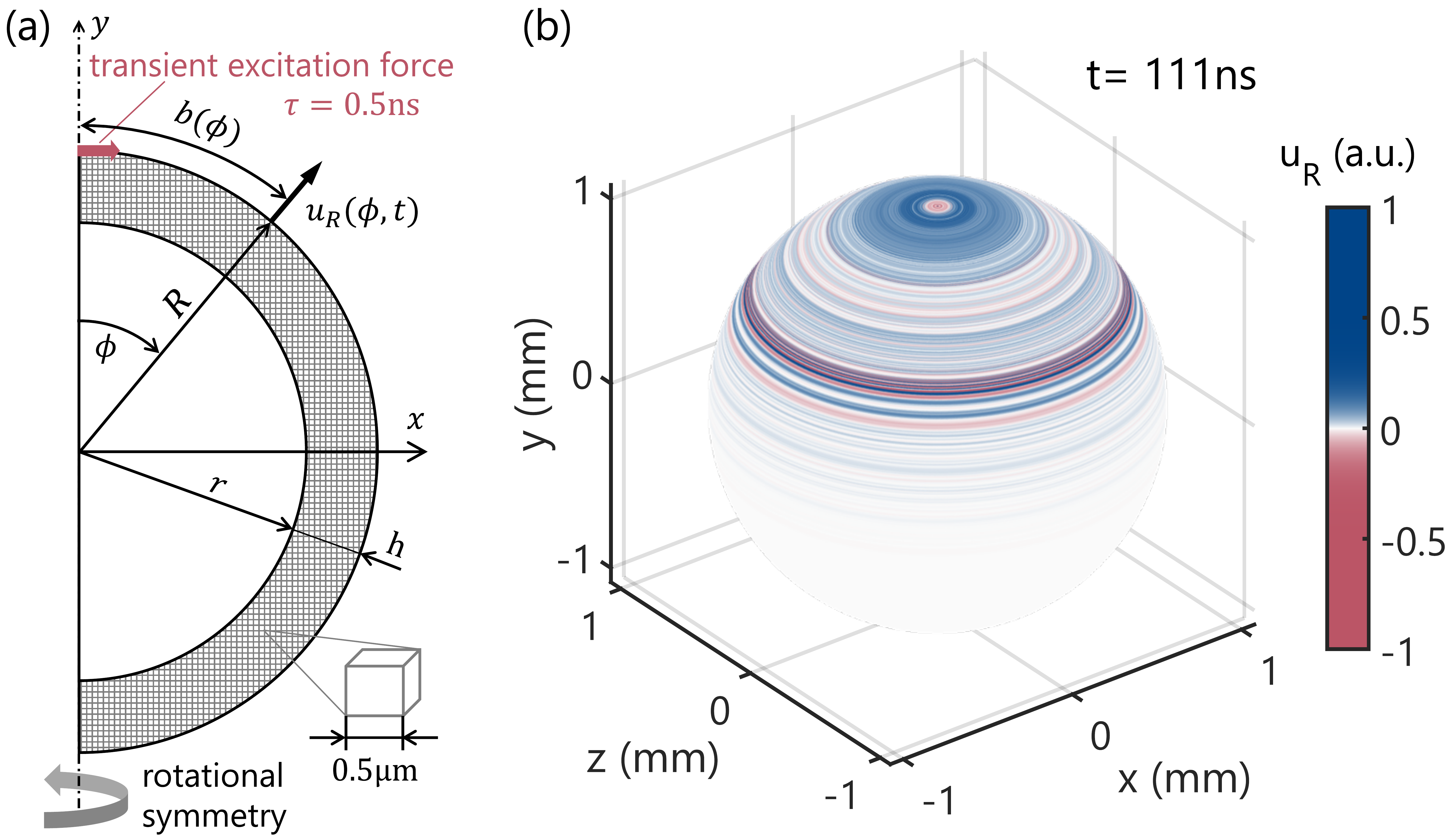}
	\caption{
        (a) Sketch of a finite element model of a HDC-capsule. 
        (b) Snapshot of the radial displacement field on \qty{80}{\micro\metre} thick HDC-capsule. 
        An animation can be found online as supplementary in \secref{sec:suppl}.
    }%
    \label{fig:FEM}
\end{figure}
In addition to the SEM plate dispersion calculations, we simulate elastic wave propagation in spherical shells using finite element method (FEM) modeling. 
We assumed isotropic elastic behavior and used the mechanical properties specified in \mytabref{tab:HDC_props}. 
Utilizing rotational symmetry along the $y$-axis, we reduced the problem to two dimensions (see \figref{fig:FEM}). 
The outer sphere radius $R$ was set to \qty{1.127}{\milli\meter} in each simulation, and the shell thickness $h$ was varied in different simulations. 
The spatial domain was discretized into a structured mesh with \qty{0.5}{\micro\metre} large cubic elements, resulting in models with around 10 million elements. 
This discretization corresponds to $83$ elements per wavelength at the target maximal frequency of \qty{250}{\mega\hertz} for the slowest wave, the Rayleigh surface acoustic wave. 
To emulate laser-based thermoelastic wave excitation, an in-plane ($x$-direction) force pulse with a spatiotemporal Gaussian profile with full width at half maximum of $\qty{2}{\micro\metre}$ and a duration of \qty{0.5}{ns} (FWHM) was applied at the north pole ($x\!=\!0$, $y\!=\!R$) of the shells. 
This approach is based on the work of Every et al.\ \cite{every_laser_2013} and was successfully applied in previous FEM elastic wave propagation studies \cite{Ryzy2018_v143_p219,grabec_surface_2022}. 
The model was solved with the commercial FEM-solver PZF\textsc{lex} (Weidlinger Associates, Mountain View, CA) using an explicit central difference time-marching scheme. 
The time-step was set to \qty{0.25}{ns} and the total simulation time to \qty{10}{\micro\second}, allowing the Rayleigh surface acoustic wave to orbit the sphere about 15 times. 
Computations were performed on a Intel Xeon W-2155 system (10 cores, \qty{3.3}{GHz}) and typically took around \qty{35}{\hour} per run. 
\Figref[(b)]{fig:FEM} shows a snapshot of the radial displacement-field $u_\mathrm{R}$ in an \qty{80}{\micro\metre} thick capsule \qty{111}{ns} after wave excitation. 
The most pronounced ring corresponds to the surface acoustic wave.

\subsection{Infrared interferometry reference measurements}
As the HDC capsule is visually opaque, interferometric thickness measurements were conducted in the infrared spectral range, using wavelengths from \qty{2}{\micro\metre} to \qty{12}{\micro\metre}, where the material is sufficiently translucent. 
General Atomics previously custom modified a compact Fourier transform infrared (FTIR) spectroscopy system to allow high precision interference modulation measurements of the outer and inner wall-surface reflections. 
The instrument is capable of measuring HDC total wall thickness and sublayer thicknesses with a \qty{5}{nm} precision \cite{caseyEvidenceThreeDimensionalAsymmetries2021, abu-shawarebLawsonCriterionIgnition2022}. 
We use this modified FTIR system to characterize an HDC capsule wall thickness variation around a great circle surrounding the equator, as a reference for FreDomLUS measurements.

\section{Results}

\subsection{Resonances on a spherical HDC shell}
\label{sec:resonances}

\begin{figure*}[tbh]
	\centering
    \includegraphics[width=\linewidth]{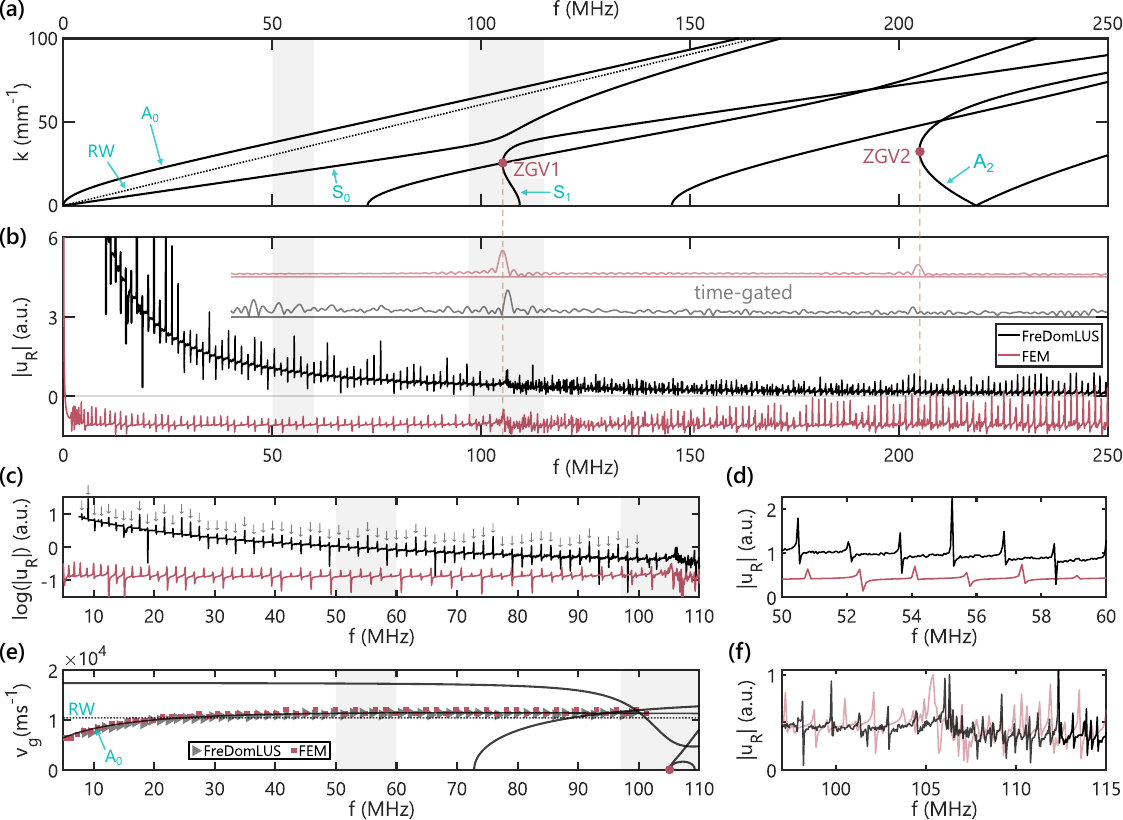}
	\caption{
        (a) Calculated elastic-wave dispersion (\texttt{GEWtool}) of a non-curved \qty{80}{\micro\metre} thick HDC-plate. 
        The fundamental \rev{anti-symmetric (A$_0$) and symmetric (S$_0$) mode, the first-order symmetric mode (S$_1$), and the second-order anti-symmetric mode (A$_2$) \cite{prada_local_2008} are} marked with an arrow, and the first two ZGV-resonances (ZGV1 \& ZGV2) with red dots. 
        The dashed line corresponds to the Rayleigh surface acoustic wave (RW) of a semi-infinite half-space of the same material. 
        (b,c,d,f) HDC-shell radial elastic response spectra from FreDomLUS-measurements (black solid lines) and FEM-simulations (red solid lines). 
        (b) Broadband response spectra up to a frequency of \qty{250}{\mega\hertz} (FEM-trace is offset by a constant value), and time-gated versions of the same traces (shifted semi-transparent lines in the upper region of the graph). 
        (c) Spectra up to \qty{110}{\mega\hertz} showing multiple fundamental mode circumferential resonance peaks. 
        Extracted resonance peak frequencies from the FreDomLUS spectrum are indicated by grey arrows. 
        (d) Zoomed view of circumferential resonance peaks (frequency range is marked as grey area in (c)). 
        (e) Calculated group velocity curves (\texttt{GEWtool}) of guided plate modes in a straight \qty{80}{\micro\metre} thick HDC-plate. 
        The grey (FreDomLUS) and black (FEM) data-points are group-velocities calculated from extracted circumferential resonance peaks. 
        (f) Zoom to the frequency region around the ZGV1-resonance (marked as grey areas in (a,b,c,e)).
    }%
    \label{fig:broadband}
\end{figure*}
\Figref[(b)]{fig:broadband} shows broadband single-point elastic response spectra recorded with a FreDomLUS microscopy setup (black solid line) and calculated from time-domain FEM simulations (red solid line) in a HDC-shell with the same dimensions ($R\!=\!\qty{1.127}{mm}$, $h\!=\!\qty{80}{\micro\metre}$). 
Both spectra are dominated by pronounced high quality factor peaks which cover the entire frequency range up to \qty{250}{\mega\hertz}. 
Except for the slowly varying background in the experiment, which can be attributed to the thermoelastic generation, the spectra match well (the FEM-spectrum is offset by a negative constant value). 
The zoomed view in \Figref[(d)]{fig:broadband} reveals a slight offset between the peaks in the measured and in the simulated spectrum. 
This slight mismatch can be expected, as the internal W-doped layer of the capsule was neglected in the FEM-model.

The low relative thickness of the shell ($\approx\!\qty{7}{\percent}$) suggests that the curvature has only a minor influence on elastic wave propagation \cite{towfighi_elastic_2003, ces_characterization_2012}. 
It is thus useful to compare its frequency spectrum to the Lamb-wave dispersion of a straight plate to spot potential ZGV-resonance peak locations. 
This is shown in the top \figref[(a)]{fig:broadband} for a \qty{80}{\micro\metre} HDC-plate. 
The lowest order ZGV-resonances (denoted as ZGV1 and ZGV2) are marked by red dots in the figure. 
Apparently, the ZGV1-resonance at around \qty{105}{\mega\hertz} aligns well with a spectral feature which clearly differs from the superimposed sharp peaks in its shape. 
However, this feature is not as pronounced and isolated as it is usually the case for ZGV-resonances in straight plates \cite{pradaLaserbasedUltrasonicGeneration2005,prada_local_2008,Gruensteidl2018_v112_p251905, ryzy_determining_2023,watzl_simultaneous_2025} (see magnified view in \figref[(f)]{fig:broadband}).

It is thus instructive to study the nature of the spectral landscape in greater detail. 
We hypothesize, that like in solid spheres \cite{shui_resonance_1988, royer_theoretical_2008}, the multiple sharp peaks are caused by constructively interfering \rev{propagating Lamb modes whose wavelength $\lambda$ corresponds to integer fractions ($n \in \mathbb{N}$) of the shell's circumference.}
The corresponding condition
\begin{equation}
    \label{eq:SAW_interference_condition}
	\rev{\lambda_n = \frac{2R\pi}{n}}
    \hspace{1ex}\Leftrightarrow\hspace{1ex} k_n\rev{:=}\frac{2\pi}{\lambda_n}=\frac{n}{R}
    \hspace{1ex}\Leftrightarrow\hspace{1ex}\Delta{}k=\frac{1}{R}
\end{equation}
provides a rule for the determination of the group velocity from neighboring resonance peaks of the same mode, as
\begin{equation}
	v_\mathrm{g}=2\pi\frac{\partial{}f}{\partial{}k}\approx2\pi\frac{\Delta{}f}{\Delta{}k}=2\pi{}R\Delta{}f . 
	\label{eq:circRes}
\end{equation}
We extract the resonance peak locations with a maximum search algorithm for frequencies lower than the observed ZGV-feature (marked by small arrows for the FreDomLUS measurement in \figref[(c)]{fig:broadband}) and used \equref{eq:circRes} for calculating the group velocity as a function of frequency. 
A comparison to the theoretical group velocity dispersion for a straight plate in \figref[(e)]{fig:broadband} shows an excellent match between experimental- (grey triangles) and simulated-data (red squares) and the fundamental anti-symmetric plate mode (black solid line denoted as A$_0$) which converges to the Rayleigh-surface acoustic wave velocity for large frequencies (not shown the graph). 
The sharp peaks will thus be denoted as circumferential resonances from here on. 
Our observation agrees well with previous findings, where we have shown that the circumferential resonances can be effectively suppressed by embedding the capsule in a soft, damping material \cite{yan_investigation_2025}. 
They further suggest that the curvature of the shell does not measurably influence wave-dispersion in the considered frequency range and thus may be neglected. 
Note that for frequencies beyond the ZGV1-point ($\gtrsim\qty{105}{\mega\hertz}$) the circumferential resonance peaks get closer spaced in frequency. 
As this behavior starts at the ZGV1-frequency we attribute it to additional circumferential resonances established by propagating Lamb-modes from the \rev{S$_0$ dispersion branch due to strong repulsion \cite{prada_local_2008, Mace2012, Veres2014crossingPoints, Gravenkamp2023} from the S$_1$ branch, whose closest approach occurs around the ZGV frequency. 
A laser-ultrasonic sensitivity analysis for the corresponding Lamb wave dispersion can be found in Fig. S3 in \secref{sec:suppl}. as in supplementary material}.

\begin{figure*}
	\centering
	\includegraphics[width=1.0\textwidth]{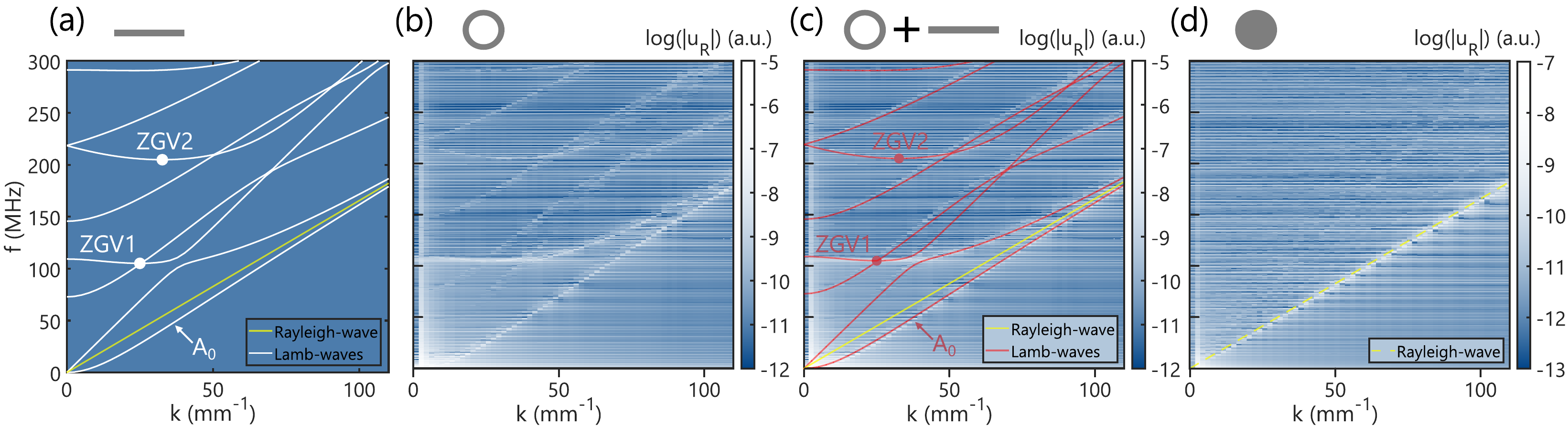} %
	\caption{
        Elastic wave dispersion curves. 
        (a) Calculated for a HDC-plate ($h\!=\!\qty{80}{\micro\metre}$) with the \texttt{GEWtool}. 
        (b,c) For a HDC shell ($R\!=\!\qty{1.127}{mm}$, $h\!=\!\qty{80}{\micro\metre}$) obtained from a FEM simulation. 
        In (c) the plate dispersion from (a) is superimposed as red family of lines. 
        (d) Dispersion of a solid HDC sphere ($R\!=\!\qty{1.127}{mm}$) obtained from a FEM simulation. 
        The $y$-axis are identical in each graph and have been omitted in (b-d) for space saving reasons.\label{fig:FEM_disp}
    }%
\end{figure*}
A more comprehensive picture about the shell wave modes can be obtained from dispersion analysis. 
For that we transformed the radial displacement field obtained from the FEM-models, $u_\mathrm{R}(b(\phi),t)$ from the physical domain to the reciprocal space to obtain $u_\mathrm{R}(k,f)$ by two-fold fast Fourier transform (FFT). 
Here, the wavenumber $k$ is the reciprocal quantity of the arc length $b\!=\!b(\phi)$ (see \figref[(a)]{fig:FEM}). 
As shown in \figref[(b)]{fig:FEM_disp}, the resulting dispersion of the shell contains the strongly non-linear modes like those of a straight plate (\figref[(a)]{fig:FEM_disp}). 
In addition, a multitude of horizontal lines appear, which stem from the circumferential resonances. 
For better comparability, plate and shell-dispersion are superimposed in (\figref[(c)]{fig:FEM_disp}). 
Here, slight deviations between the Lamb-like modes in the shell and those of the straight plate are revealed. 
We attribute this to the curvature of the shell. 
But most importantly, the first two ZGV-points clearly exist in the shell's dispersion. 
Thus, we conclude that observed spectral feature in measurements and simulations around \qty{105}{\mega\hertz} (see \figref[(b,f)]{fig:broadband}) is caused by a ZGV-resonance. 
For sanity checking, we additionally conducted FEM simulations in a solid sphere with the same size as the capsule. 
As expected, its dispersion (\figref[(d)]{fig:FEM_disp}) contains horizontal circumferential modes, but not the non-linear plate modes. 
Here, the circumferential resonance are created by a non-dispersive SAW which appears as pronounced straight line in the dispersion and matches well with the Rayleigh-SAW of a semi-infinite solid. 
We conclude from the dispersion analysis, that the shell's dispersion landscape may be considered as a superposition of a solid sphere's dispersion and the dispersion of a straight plate, in that sense, that both, plate-resonances and circumferential resonances appear.

Having revealed the nature of the observed shell resonances, we can isolate the ZGV-resonance from the superimposed circumferential resonances by employing their different build-up times. 
While ZGV-resonances can be considered as constructively interfering bulk waves, being back and forth reflected from the quasi-parallel shell surfaces \cite{rose_ultrasonic_2004}, the circumferential resonances originate from slower propagating guided waves (the fundamental mode A$_{0}$) which travel a larger distance around the circumference before interfering which one another. 
Hence, they require more time to build-up, and we can suppress them by selecting only early contributions in time-domain signals ('time-gating'). 
We therefore inverse fast Fourier transform (iFFT) acoustic spectra to the time domain, and back-transform only signals earlier than a certain gate-time. 
This has been done for the broadband spectra shown \figref[(b)]{fig:broadband} by selecting time domain signal with $t\!<\!\qty{0.55}{\micro\second}$. 
In both resulting spectra (denoted as 'time-gated') the sharp circumferential resonance peaks have vanished, while the ZGV1-resonance remains. 
In addition, a second ZGV2-resonance has appeared, which was not visible in the unprocessed spectra. 
Note that while both ZGV-resonance peaks in the FEM spectrum perfectly match the plate dispersion predictions (\figref[(a)]{fig:broadband}), the experimental resonances are slightly off. 
As the center-frequency ratio of two independent resonances in a homogeneous plate (like the observed the ZGV1 and ZGV2 resonances) is a unique function of Poisson's ratio \cite{clorennec_local_2007,watzl_situ_2022}, we can estimate Poisson's ratio by using a numerically calculated calibration curve (see for example Fig.\,4 in reference \cite{clorennec_local_2007}). 
\rev{From the measurement in \figref[(b)]{fig:broadband}, we could extracted the frequency values of the first two ZGV resonances, which are $f_\mathrm{ZGV1}$ = \qty{106.34}{\mega\hertz} and $f_\mathrm{ZGV2}$ = \qty{203.28}{\mega\hertz}, respectively.}
With \rev{the experimental} ratio $\nicefrac{f_\mathrm{ZGV2}}{f_\mathrm{ZGV1}}\!\approx\!1.91$, we obtain an effective Poisson's ratio of $\nu\!\approx\!0.16$ instead of \rev{the reference value} $\nu\!=\!0.1$ as given in \mytabref{tab:HDC_props}.
This deviation may be to some extent caused by the slight W-doping of the HDC-capsule.
\rev{However, it should be noted that a wide range of Poisson's ratio values of diamond-like carbon are reported in literature spanning from $\nu=\numrange[range-phrase=\text{~to~}]{0.04}{0.2}$ \cite{Klein1993diamond, Cho2005diamond}, and our obtained value falls within the upper end of this range.} 

\subsection{ZGV-resonance wall thickness measurements}\label{sec:mapping}
\begin{figure*}
	\centering
	\includegraphics[width=1.0\linewidth]{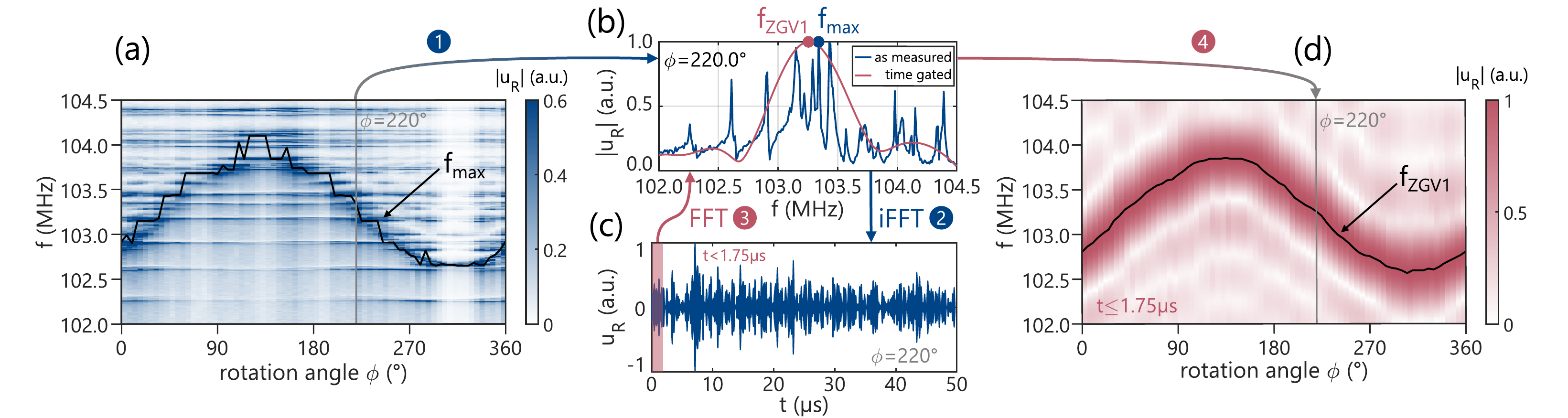}
	\caption{
        (a) Narrow-band FreDomLUS-measured acoustic response spectra in the range of the ZGV1-resonance, recorded along the HDC-shell equator. 
        The spectra are normalized to their maximum magnitudes (black solid line). 
        (b) Example spectrum recorded at a rotation angle of $\phi\!=\!\qty{220}{\degree}$. 
        (c) Time-signal of the spectrum in (b), obtained by inverse fast Fourier transformation. 
        The shaded red area indicates the fraction of the signal which is selected prior back-transformation. 
        The time-gated spectrum is plotted as red solid line in (b). 
        (d) Time-gated acoustic response spectra as function of rotation angle $\phi$. 
        The black solid line indicates the extracted ZGV1-resonance frequency. 
        The sequence of the applied data-processing steps is indicated by the  numbers \ding{202}\,--\,\ding{205}.\label{fig:phiScan}
    }%
\end{figure*}
For mapping the HDC-shell wall thickness, we recorded narrow-band acoustic spectra around the ZGV1-frequency with the FreDomLUS microscopy system along the equator of the HDC-shell. 
Therefore, the shell was rotated in $\Delta\phi\!=\!\qty{5}{\degree}$ steps. 
The resulting spectra are shown in a 2D-map in \figref[(a)]{fig:phiScan}. 
In addition to position independent circumferential resonance-peaks, which emerge as horizontal lines in the plot, the map clearly shows the varying ZGV1-resonance feature, linked to a wall thickness change. 
\Figref[(b)]{fig:phiScan} depicts an example spectrum, recorded at an rotation angle of \qty{220}{\degree}. 
Here, the ZGV-resonance is clearly superimposed by circumferential resonances, and by simply selecting the maximum peak frequency $f_\mathrm{max}$ (blue dot in \Figref[(b)]{fig:phiScan}), only a rough approximation the ZGV1-resonance frequency can be obtained. 
This is vividly illustrated in \Figref[(a)]{fig:phiScan}, where the black $f_\mathrm{max}$-line always snaps to the nearest circumferential resonance position. 
We thus point out that for getting precise wall thickness results, it is necessary to isolate the ZGV-resonance from the circumferential resonances by the time-gating procedure as described before. 
Here, we choose a cut-off time of \qty{1.75}{\micro\second} (\figref[(c)]{fig:phiScan}) which corresponds to around 2.8 round-trips of the SAW (A$_0$-mode, with a maximum group velocity of \qty{11414}{\metre\per\second}), and to 127 back and forth reflection of the transverse bulk wave. 
This suppresses the circumferential resonances in the processed spectrum, while the ZGV1-resonance remains (red line in \figref[(b)]{fig:phiScan}. 
Note that the ZGV1-peak is artificially broadened due to the use of a finite time-window. 
However, this does not alter the location of the ZGV1-peak maximum $f_\mathrm{ZGV1}$. 
The latter is clearly shifted compared to the unprocessed spectrum. 
\rev{For the selection of the time-window size, a brief analysis is discussed in Fig. S2 in supplementary (Sec. \ref{sec:suppl}); and for more details of time-gating method, readers are referred to the Suppl. Mat. of Ref. \cite{grunsteidl_measurement_2020}, which will not be repeated here.}
Finally, in the processed 2D-map (\figref[(d)]{fig:phiScan}) the circumferential resonances have vanished, while the varying ZGV1-peak remains, and the extracted ZGV1 center peak progression is substantially smoother than in the unprocessed map (compare black $f_\mathrm{max}$ and $f_\mathrm{ZGV1}$ curves in (a) and (d)).

\begin{figure}
	\centering
	\includegraphics[width=0.45\textwidth]{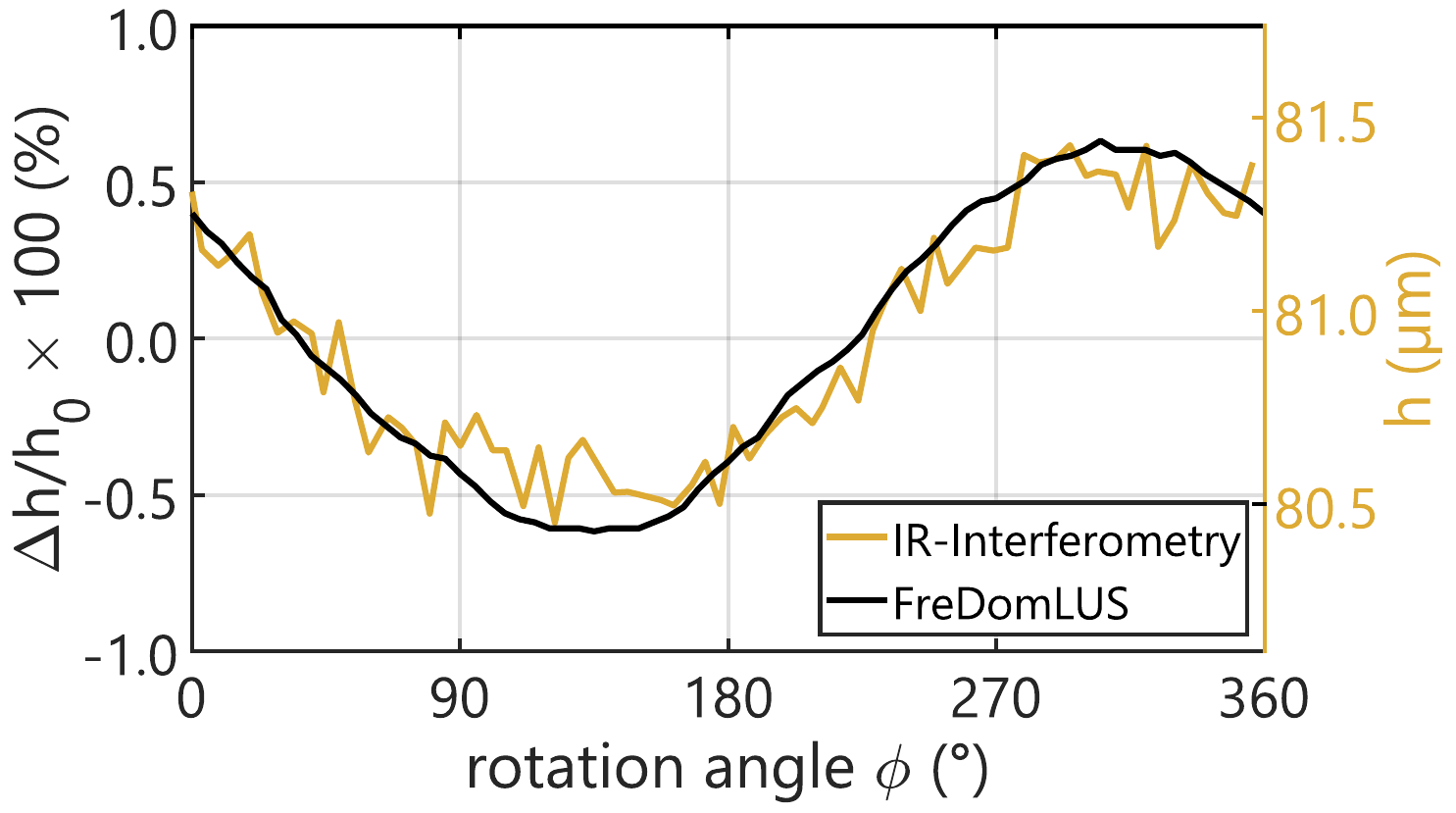}
	\caption{
        Measured relative (left $y$-axis) and absolute (right $y$-axis) wall thickness of a HDC-shell with an average diameter of \qty{2.254}{mm}. 
        Calculated from ZGV1-resonance center frequencies obtained from FreDomLUS-measurements (black solid line) and measured by IR-interferometry (yellow solid line).\label{fig:result}
    }%
\end{figure}
As shown in \secref{sec:resonances}, the guided modes in a curved shell behave like Lamb waves in the relevant frequency range. 
Thus, we use the inverse plate-thickness scaling law (\equref{eq:ZGV_scaling}) to calculate the relative wall thickness change along the equator of the HDC-shell via (see Appendix\,\ref{sec:validation} for further validation). 
\begin{equation}
	\frac{\Delta{}h}{h_0} = \frac{h(\phi)-h_0}{h_0} = -\frac{f_\mathrm{ZGV1}(\phi)-f_\mathrm{ZGV1,0}}{f_\mathrm{ZGV1,0}}. 
\label{eq:wall_thickness}
\end{equation}
Here, we use the average ZGV1-frequency $\langle{}f_\mathrm{ZGV1}\rangle_\phi\!=\!\qty{103.22}{\mega\hertz}$ as reference frequency $f_\mathrm{ZGV1,0}$. 
The resulting black solid curve in \figref{fig:result} reveals a smooth wall thickness variation by a little more than \qty{1}{\percent} along the capsule equator. 
To enable comparison to IR-interferometry reference wall thickness measurements, which provide an absolute measure (yellow line in \figref{fig:result}, right hand side $y$-axis), we used its average result as a reference $h_0\!=\!\qty{80.93}{\micro\metre}$ to get $h(\phi)$ from \equref{eq:wall_thickness}. 
As shown in the plot, the results agree extraordinarily well and show a wall thickness variation of about \qty{1}{\micro\metre} along the equator. 
Note that the FreDomLUS-measurement provides a smoother curve-progression than the reference measurement. 

\rev{To check whether this is caused by a spatial averaging effect, we calculate the lateral extent of the resonance used, which provides an approximate measure of the spatial resolution of our method --- provided that the laser spot sizes are sufficiently small. } 
It can be estimated by half of the resonance wavelength \cite{prada_local_2008, ces_edge_2011}, given by $0.5\lambda_\mathrm{ZGV1}\!=\!\pi{}k_\mathrm{ZGV1}^{-1}\!=\!\qty{125.7}{\micro\metre}$ (with \rev{a numerically calculated value of} $k_\mathrm{ZGV1}\!=\!\qty{25.0}{mm^{-1}}$; see dispersion calculations in \figref[(a)]{fig:broadband}). 
By using the measured capsule radius \rev{($R\!=\!\qty{1.127}{\milli\meter}$)}, this corresponds to an angular resolution of $\Delta{}\phi\!=\!\qty{6.4}{\degree}$. 
As this is in the range of the \rev{angular step size} (\qty{5.0}{\degree} for FreDomLUS, and \qty{5.1}{\degree} for IR-interferometry), we assume that averaging effects can be neglected, and the FreDomLUS wall thickness measurement is more precise than the IR-reference method.

\section{Discussion and Conclusion}\label{sec:conclusion}
We have presented an non-contact, local wall thickness variation measurement method for hollow spherical inertial confinement fusion targets based on a frequency domain laser ultrasound microscopy technique. 
It relies on the inverse scaling of the ZGV-resonance frequencies with the shell-thickness. 
For demonstration purposes, we scanned the first ZGV-resonance at around $\qty{106}{\mega\hertz}$ along the equator of a \qty{2.154}{mm}-diameter W-doped HDC capsule with an average wall thickness of $\qty{81}{\micro\metre}$ (relative wall thickness of \rev{$\qty{7}{\percent}$} with respect to its radius). 
We found wall thickness variations of around $\qty{1}{\percent}$ and an excellent match to IR-interferometry reference measurements.

To justify our approach, relying on the applicability of Lamb's theory for infinite, non-curved plates, we carefully studied the inherently twofold-resonant nature of spherical shells. 
For that purpose we combined plate-wave dispersion calculations, FEM wave propagation simulations and experimental observations. 
We showed that in the considered frequency range from about \qty{5}{\mega\hertz} to \qty{250}{\mega\hertz} and for the specific normalized curvature (relative shell thickness) of $\nicefrac{h}{R}\!\approx\!\qty{7}{\percent}$, the shell's dispersion comprises of non-linear branches that closely match those of a straight plate. 
In addition, it contains circumferential resonances which are a consequence of the closed-loop surface that sets an additional resonance condition for propagating Lamb-modes orbiting the shell. 
These resonances emerge as multiple high quality factor peaks in broadband acoustic spectra that obscure the required ZGV-resonances. 
We demonstrated that the different build up times of the two resonance species can be used to effectively isolate ZGV-resonances by selecting early contributions in time-domain signals.

A second ZGV-resonance at $\approx\qty{203}{\mega\hertz}$ was revealed after signal-processing and has been used to estimate the Poisson's ratio of $\nu\!=\!0.16$, substantially deviating from the HDC value of \num{0.10} \rev{reported for the diamond base material by the supplier \cite{noauthor_diamond_2025}}. 
This indicates a modulation of the elastic parameters by the internal W-doped layer. 
In future work, we plan to assess the potential of using the circumferential resonances from single-point, broadband spectra for elastic characterization. 
For instance, by extracting the circumferential resonance peaks up to the first ZGV-resonance, which were shown to belong to the anti-symmetric fundamental mode (A$_0$) and can be used to calculate it's group velocity. 
Simultaneously fitting the A$_0$ group velocity and the two observed ZGV-resonance frequencies with a fast plate-dispersion forward model may be used to determine the elastic constants and the average wall thickness of the shell, provided that the shell's outer sphere surface is perfect.

Since for sufficiently large frequencies the A$_0$-mode can be considered as a surface acoustic wave propagating on the outer shell boundary, it must be affected by its geometry. 
Thus, we hypothesize that the A$_0$-mode circumferential resonance quality factors may be used to detect deviations from a perfect spherical geometry.

The lateral resolution of the wall thickness measurement method is \rev{primarily determined by the spatial extent of the employed ZGV-resonance. It} can be approximated by half of the resonance wavelength\rev{, which for the ZGV1 mode corresponds to about \qty{126}{\micro\metre}. Using the ZGV2 resonance, which has a shorter wavelength, this resolution could} potentially be improved by about \qty{20}{\percent}.

Finally, we point out that the method can be applied to optically opaque materials like metals, or metal-doped HDC, where optical wall thickness characterization methods are limited due to low penetration depths.

\appendix

\printcredits

\section*{Acknowledgements}\label{}
This collaborative work is supported by General Atomics Internal R\&D fund and is co-financed by the project “HyperMAT” by the federal government of Upper Austria and the European Regional Development Fund (EFRE) in the framework of the EU-program IBW/EFRE \& JTF 2021-2027. 
The spherical HDC coating were produced by Diamond Materials GmbH, General Atomics laser drilled, leached the article into a capsule, built into Capsule-Fill-Tube-Assembly (CFTA), then into fully assembled targets for National Ignition Facility (NIF) shots under US Department of Energy (DOE) contract No.\ 89233124CNA000365. 
The authors would like to thank Marc Choquet (Tecnar) for initiating the collaboration between GA and RECENDT, and Robert Holzer (RECENDT) for taking a high-quality photo of our sample. 
For all figures, we used a high-contrast color-scheme which is suitable for people with impaired color vision as proposed by Paul Tol (\url{https://sronpersonalpages.nl/~pault}).

\section*{Declaration of competing interest}
The authors declare that they have no competing financial interests or personal relationships that could have appeared to influence the work reported in this paper.

\section*{Data availability}
Data will be made available by the authors upon reasonable request.

\section{Supplementary data}\label{sec:suppl}
A supplementary video \rev{to this article can be found online as supplementary material; along with a brief analysis of the impact of the thin W-layer on the ZGV resonance frequency, a justification for the choice of time-gating, as well as a corresponding Lamb wave LUS sensitivity calculation}.

\section{Validation \rev{of} ZGV-frequency scaling law}
\label{sec:validation}

\begin{figure}
	\centering
	\includegraphics[scale=0.85]{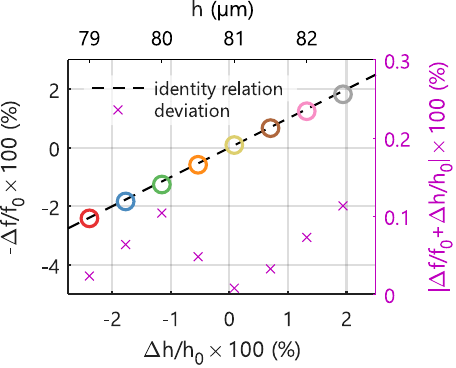}
	\caption{
        Validation of inverse scaling of the ZGV1-resonance frequency with shell-thickness with FEM simulations. 
        The colored data-points correspond to ZGV1-resonance frequencies extracted from simulated broadband spectra of HDC-capsules with corresponding thickness $h$ (see time-gated FEM-trace in \figref[(b)]{fig:broadband} for instance).
        \rev{The purple 'x' symbols with the corresponding second y-axis on the right represent the relative deviation of the frequency progression from the identity relation (i.\,e.\ a one-to-one correlation between frequency and thickness change).}
    }%
    \label{fig:calib}
\end{figure}

To confirm the validity of \equref{eq:ZGV_scaling} we performed a series of FEM-simulations where we varied the shell-thickness $h$ in $\qty{0.5}{\micro\metre}$ steps from \rev{\SIrange{79}{82.5}{\micro\meter}}. 
From time-gated spectra (as shown in \figref[(b)]{fig:broadband}) we extracted the ZGV1-frequency by maximum search. 
\Figref{fig:calib} shows that the negative relative change of the ZGV1-resonance frequency ($-\nicefrac{\Delta{}f}{f}$) linearly scales with the relative change of shell-thickness ($\nicefrac{\Delta{}h}{h}$) \rev{in a one-to-one relationship within a margin of about \qty{0.1}{\percent} for the simulated range of deviations}. 
Thus, \equref{eq:ZGV_scaling} approximately holds for the studied range of wall thickness variations.

\bibliographystyle{Refs-style-model3a-num-names}
\bibliography{Refs-HDC-capsule}

\begin{thebibliography}{65}
\providecommand{\natexlab}[1]{#1}
\providecommand{\url}[1]{\texttt{#1}}
\providecommand{\href}[2]{#2}
\providecommand{\path}[1]{#1}
\providecommand{\DOIprefix}{doi:}
\providecommand{\ArXivprefix}{arXiv:}
\providecommand{\URLprefix}{URL: }
\providecommand{\Pubmedprefix}{pmid:}
\providecommand{\doi}[1]{\href{http://dx.doi.org/#1}{\path{#1}}}
\providecommand{\Pubmed}[1]{\href{pmid:#1}{\path{#1}}}
\providecommand{\bibinfo}[2]{#2}
\ifx\xfnm\undefined \def\xfnm[#1]{\unskip,\space#1}\fi
\makeatletter\def\@biblabel#1{#1.}\makeatother
\bibitem[{Hurricane et~al.(2023)Hurricane, Patel, Betti, Froula, Regan, Slutz et~al.}]{hurricane_physics_2023}
\bibinfo{author}{Hurricane\xfnm[ O.A.]}, \bibinfo{author}{Patel\xfnm[ P.K.]}, \bibinfo{author}{Betti\xfnm[ R.]}, \bibinfo{author}{Froula\xfnm[ D.H.]}, \bibinfo{author}{Regan\xfnm[ S.P.]}, \bibinfo{author}{Slutz\xfnm[ S.A.]}, et~al.
\newblock \bibinfo{title}{Physics principles of inertial confinement fusion and {{U}}.{{S}}. program overview}.
\newblock \emph{\bibinfo{journal}{Reviews of Modern Physics}} \bibinfo{year}{2023};\hspace{0pt}\textbf{\bibinfo{volume}{95}}(\bibinfo{number}{2}):\bibinfo{pages}{025005}.
\newblock \DOIprefix\doi{10.1103/RevModPhys.95.025005}.
\bibitem[{{Abu-Shawareb} et~al.(2024){Abu-Shawareb}, Acree, Adams, Adams, Addis, Aden et~al.}]{abu-shawareb_achievement_2024}
\bibinfo{author}{{Abu-Shawareb}\xfnm[ H.]}, \bibinfo{author}{Acree\xfnm[ R.]}, \bibinfo{author}{Adams\xfnm[ P.]}, \bibinfo{author}{Adams\xfnm[ J.]}, \bibinfo{author}{Addis\xfnm[ B.]}, \bibinfo{author}{Aden\xfnm[ R.]}, et~al.
\newblock \bibinfo{title}{Achievement of {{Target Gain Larger}} than {{Unity}} in an {{Inertial Fusion Experiment}}}.
\newblock \emph{\bibinfo{journal}{Physical Review Letters}} \bibinfo{year}{2024};\hspace{0pt}\textbf{\bibinfo{volume}{132}}(\bibinfo{number}{6}):\bibinfo{pages}{065102}.
\newblock \DOIprefix\doi{10.1103/PhysRevLett.132.065102}.
\bibitem[{Biener et~al.(2009)Biener, Ho, Wild, Woerner, Biener, {El-dasher} et~al.}]{biener_diamond_2009}
\bibinfo{author}{Biener\xfnm[ J.]}, \bibinfo{author}{Ho\xfnm[ D.]}, \bibinfo{author}{Wild\xfnm[ C.]}, \bibinfo{author}{Woerner\xfnm[ E.]}, \bibinfo{author}{Biener\xfnm[ M.]}, \bibinfo{author}{{El-dasher}\xfnm[ B.]}, et~al.
\newblock \bibinfo{title}{Diamond spheres for inertial confinement fusion}.
\newblock \emph{\bibinfo{journal}{Nuclear Fusion}} \bibinfo{year}{2009};\hspace{0pt}\textbf{\bibinfo{volume}{49}}(\bibinfo{number}{11}):\bibinfo{pages}{112001}.
\newblock \DOIprefix\doi{10.1088/0029-5515/49/11/112001}.
\bibitem[{Ross et~al.(2015)Ross, Ho, Milovich, D{\"o}ppner, McNaney, MacPhee et~al.}]{ross_high-density_2015}
\bibinfo{author}{Ross\xfnm[ J.S.]}, \bibinfo{author}{Ho\xfnm[ D.]}, \bibinfo{author}{Milovich\xfnm[ J.]}, \bibinfo{author}{D{\"o}ppner\xfnm[ T.]}, \bibinfo{author}{McNaney\xfnm[ J.]}, \bibinfo{author}{MacPhee\xfnm[ A.G.]}, et~al.
\newblock \bibinfo{title}{High-density carbon capsule experiments on the national ignition facility}.
\newblock \emph{\bibinfo{journal}{Physical Review E}} \bibinfo{year}{2015};\hspace{0pt}\textbf{\bibinfo{volume}{91}}(\bibinfo{number}{2}):\bibinfo{pages}{021101}.
\newblock \DOIprefix\doi{10.1103/PhysRevE.91.021101}.
\bibitem[{Cardenas et~al.(2018)Cardenas, Schmidt, Loomis, Randolph, Hamilton, Oertel et~al.}]{cardenas_progress_2018}
\bibinfo{author}{Cardenas\xfnm[ T.]}, \bibinfo{author}{Schmidt\xfnm[ D.W.]}, \bibinfo{author}{Loomis\xfnm[ E.N.]}, \bibinfo{author}{Randolph\xfnm[ R.B.]}, \bibinfo{author}{Hamilton\xfnm[ C.E.]}, \bibinfo{author}{Oertel\xfnm[ J.]}, et~al.
\newblock \bibinfo{title}{Progress {{Toward Fabrication}} of {{Machined Metal Shells}} for the {{First Double-Shell Implosions}} at the {{National Ignition Facility}}}.
\newblock \emph{\bibinfo{journal}{Fusion Science and Technology}} \bibinfo{year}{2018};\hspace{0pt}\textbf{\bibinfo{volume}{73}}(\bibinfo{number}{3}):\bibinfo{pages}{344--353}.
\newblock \DOIprefix\doi{10.1080/15361055.2017.1406251}.
\bibitem[{Braun et~al.(2023)Braun, Kucheyev, Shin, Wang, Ye, Teslich~Jr et~al.}]{braun_tungsten_2023}
\bibinfo{author}{Braun\xfnm[ T.]}, \bibinfo{author}{Kucheyev\xfnm[ S.]}, \bibinfo{author}{Shin\xfnm[ S.]}, \bibinfo{author}{Wang\xfnm[ Y.]}, \bibinfo{author}{Ye\xfnm[ J.]}, \bibinfo{author}{Teslich~Jr\xfnm[ N.]}, et~al.
\newblock \bibinfo{title}{Tungsten doped diamond shells for record neutron yield inertial confinement fusion experiments at the {{National Ignition Facility}}}.
\newblock \emph{\bibinfo{journal}{Nuclear Fusion}} \bibinfo{year}{2023};\hspace{0pt}\textbf{\bibinfo{volume}{63}}(\bibinfo{number}{1}):\bibinfo{pages}{016022}.
\newblock \DOIprefix\doi{10.1088/1741-4326/aca4e4}.
\bibitem[{Kritcher et~al.(2024)Kritcher, Schlossberg, Weber, Young, Dewald, Zylstra et~al.}]{kritcher_design_2024}
\bibinfo{author}{Kritcher\xfnm[ A.L.]}, \bibinfo{author}{Schlossberg\xfnm[ D.]}, \bibinfo{author}{Weber\xfnm[ C.]}, \bibinfo{author}{Young\xfnm[ C.]}, \bibinfo{author}{Dewald\xfnm[ E.]}, \bibinfo{author}{Zylstra\xfnm[ A.]}, et~al.
\newblock \bibinfo{title}{Design of first experiment to achieve fusion target gain {$>$} 1}.
\newblock In: \bibinfo{editor}{Phipps\xfnm[ C.R.]}, \bibinfo{editor}{Gruzdev\xfnm[ V.E.]}, editors. \emph{\bibinfo{booktitle}{High-{{Power Laser Ablation VIII}}}}. \bibinfo{address}{Santa Fe, United States}: \bibinfo{publisher}{SPIE}.
\newblock ISBN \bibinfo{isbn}{978-1-5106-7184-3 978-1-5106-7185-0}; \bibinfo{year}{2024}, \hspace{0pt}p.~\bibinfo{pages}{1}.
\newblock \DOIprefix\doi{10.1117/12.3017211}.
\bibitem[{Tollefson(2024)}]{tollefson_how_2024}
\bibinfo{author}{Tollefson\xfnm[ J.]}.
\newblock \bibinfo{title}{How the world's biggest laser smashed a nuclear-fusion record}.
\newblock \emph{\bibinfo{journal}{online}} \bibinfo{year}{2024};\hspace{0pt}.
\bibitem[{Casey et~al.(2021)Casey, MacGowan, Sater, Zylstra, Landen, Milovich et~al.}]{caseyEvidenceThreeDimensionalAsymmetries2021}
\bibinfo{author}{Casey\xfnm[ D.T.]}, \bibinfo{author}{MacGowan\xfnm[ B.J.]}, \bibinfo{author}{Sater\xfnm[ J.D.]}, \bibinfo{author}{Zylstra\xfnm[ A.B.]}, \bibinfo{author}{Landen\xfnm[ O.L.]}, \bibinfo{author}{Milovich\xfnm[ J.]}, et~al.
\newblock \bibinfo{title}{Evidence of {{Three-Dimensional Asymmetries Seeded}} by {{High-Density Carbon-Ablator Nonuniformity}} in {{Experiments}} at the {{National Ignition Facility}}}.
\newblock \emph{\bibinfo{journal}{Physical Review Letters}} \bibinfo{year}{2021};\hspace{0pt}\textbf{\bibinfo{volume}{126}}(\bibinfo{number}{2}):\bibinfo{pages}{025002}.
\newblock \DOIprefix\doi{10.1103/PhysRevLett.126.025002}.
\bibitem[{Scruby and Drain(1990{\natexlab{a}})}]{scrubyLaserUltrasonicsTechniques1990}
\bibinfo{author}{Scruby\xfnm[ C.B.]}, \bibinfo{author}{Drain\xfnm[ L.E.]}.
\newblock \emph{\bibinfo{title}{Laser {{Ultrasonics Techniques}} and {{Applications}}}}.
\newblock \bibinfo{publisher}{Adam Hilger}; \bibinfo{year}{1990}{\natexlab{a}}.
\newblock ISBN \bibinfo{isbn}{0-7503-0050-7}.
\bibitem[{Cheeke(2012)}]{cheeke_nondestructive_2012}
\bibinfo{author}{Cheeke\xfnm[ J.D.N.]}.
\newblock \bibinfo{title}{Nondestructive {{Evaluation}} of {{Materials}} - {{Thickness Gauging}}}.
\newblock In: \emph{\bibinfo{booktitle}{Fundamentals and {{Applications}} of {{Ultrasonic Waves}}}}. \bibinfo{publisher}{CRC Press}; \bibinfo{year}{2012}, \hspace{0pt}p. \bibinfo{pages}{402--407}.
\bibitem[{Viktorov(1967)}]{viktorov_rayleigh_1967}
\bibinfo{author}{Viktorov\xfnm[ I.A.]}.
\newblock \emph{\bibinfo{title}{Rayleigh and {{Lamb Waves}}}}.
\newblock Ultrasonic {{Technology}}. \bibinfo{publisher}{Springer Science + Business Media, LCC}; \bibinfo{edition}{1} ed.; \bibinfo{year}{1967}.
\newblock ISBN \bibinfo{isbn}{978-1-4899-5683-5}.
\bibitem[{Graff(1991)}]{graff1991_p431}
\bibinfo{author}{Graff\xfnm[ K.F.]}.
\newblock \bibinfo{title}{8.1 {{Continuous}} waves in a plate}.
\newblock In: \emph{\bibinfo{booktitle}{Wave Motion in Elastic Solids}}. \bibinfo{address}{New York}: \bibinfo{publisher}{Dover Publications Inc.}
\newblock ISBN \bibinfo{isbn}{0-486-66745-6}; \bibinfo{year}{1991}, \hspace{0pt}p. \bibinfo{pages}{431--463}.
\bibitem[{Tolstoy and Usdin(1957)}]{tolstoy_wave_1957}
\bibinfo{author}{Tolstoy\xfnm[ I.]}, \bibinfo{author}{Usdin\xfnm[ E.]}.
\newblock \bibinfo{title}{Wave {{Propagation}} in {{Elastic Plates}}: {{Low}} and {{High Mode Dispersion}}}.
\newblock \emph{\bibinfo{journal}{The Journal of the Acoustical Society of America}} \bibinfo{year}{1957};\hspace{0pt}\textbf{\bibinfo{volume}{29}}(\bibinfo{number}{1}):\bibinfo{pages}{37--42}.
\newblock \DOIprefix\doi{10.1121/1.1908675}.
\bibitem[{Gibson and Popovics(2005)}]{gibson_lamb_2005}
\bibinfo{author}{Gibson\xfnm[ A.]}, \bibinfo{author}{Popovics\xfnm[ J.S.]}.
\newblock \bibinfo{title}{Lamb {{Wave Basis}} for {{Impact-Echo Method Analysis}}}.
\newblock \emph{\bibinfo{journal}{Journal of Engineering Mechanics}} \bibinfo{year}{2005};\hspace{0pt}\textbf{\bibinfo{volume}{131}}(\bibinfo{number}{4}):\bibinfo{pages}{438--443}.
\newblock \DOIprefix\doi{10.1061/(ASCE)0733-9399(2005)131:4(438)}.
\bibitem[{Prada et~al.(2005)Prada, Balogun and Murray}]{pradaLaserbasedUltrasonicGeneration2005}
\bibinfo{author}{Prada\xfnm[ C.]}, \bibinfo{author}{Balogun\xfnm[ O.]}, \bibinfo{author}{Murray\xfnm[ T.W.]}.
\newblock \bibinfo{title}{Laser-based ultrasonic generation and detection of zero-group velocity {{Lamb}} waves in thin plates}.
\newblock \emph{\bibinfo{journal}{Applied Physics Letters}} \bibinfo{year}{2005};\hspace{0pt}\textbf{\bibinfo{volume}{87}}(\bibinfo{number}{19}):\bibinfo{pages}{194109}.
\newblock \DOIprefix\doi{10.1063/1.2128063}.
\bibitem[{Prada et~al.(2008{\natexlab{a}})Prada, Clorennec and Royer}]{prada_local_2008}
\bibinfo{author}{Prada\xfnm[ C.]}, \bibinfo{author}{Clorennec\xfnm[ D.]}, \bibinfo{author}{Royer\xfnm[ D.]}.
\newblock \bibinfo{title}{Local vibration of an elastic plate and zero-group velocity {{Lamb}} modes}.
\newblock \emph{\bibinfo{journal}{The Journal of the Acoustical Society of America}} \bibinfo{year}{2008}{\natexlab{a}};\hspace{0pt}\textbf{\bibinfo{volume}{124}}(\bibinfo{number}{1}):\bibinfo{pages}{203--212}.
\newblock \DOIprefix\doi{10.1121/1.2918543}.
\bibitem[{Every(2016)}]{every_intersections_2016}
\bibinfo{author}{Every\xfnm[ A.G.]}.
\newblock \bibinfo{title}{Intersections of the {{Lamb}} mode dispersion curves of free isotropic plates}.
\newblock \emph{\bibinfo{journal}{The Journal of the Acoustical Society of America}} \bibinfo{year}{2016};\hspace{0pt}\textbf{\bibinfo{volume}{139}}(\bibinfo{number}{4}):\bibinfo{pages}{1793--1798}.
\newblock \DOIprefix\doi{10.1121/1.4946771}.
\bibitem[{Spytek et~al.(2023)Spytek, Ambrozinski and Pelivanov}]{spytek_non-contact_2023}
\bibinfo{author}{Spytek\xfnm[ J.]}, \bibinfo{author}{Ambrozinski\xfnm[ L.]}, \bibinfo{author}{Pelivanov\xfnm[ I.]}.
\newblock \bibinfo{title}{Non-contact detection of ultrasound with light -- {{Review}} of recent progress}.
\newblock \emph{\bibinfo{journal}{Photoacoustics}} \bibinfo{year}{2023};\hspace{0pt}\textbf{\bibinfo{volume}{29}}:\bibinfo{pages}{100440}.
\newblock \DOIprefix\doi{10.1016/j.pacs.2022.100440}.
\bibitem[{Watzl et~al.(2025)Watzl, Ryzy, {\"O}sterreicher, Arnoldt, Yan, Scherleitner et~al.}]{watzl_simultaneous_2025}
\bibinfo{author}{Watzl\xfnm[ G.]}, \bibinfo{author}{Ryzy\xfnm[ M.]}, \bibinfo{author}{{\"O}sterreicher\xfnm[ J.A.]}, \bibinfo{author}{Arnoldt\xfnm[ A.R.]}, \bibinfo{author}{Yan\xfnm[ G.]}, \bibinfo{author}{Scherleitner\xfnm[ E.]}, et~al.
\newblock \bibinfo{title}{Simultaneous laser ultrasonic measurement of sound velocities and thickness of plates using combined mode local acoustic spectroscopy}.
\newblock \emph{\bibinfo{journal}{Ultrasonics}} \bibinfo{year}{2025};\hspace{0pt}\textbf{\bibinfo{volume}{145}}:\bibinfo{pages}{107453}.
\newblock \DOIprefix\doi{10.1016/j.ultras.2024.107453}.
\bibitem[{Gr{\"u}nsteidl et~al.(2018)Gr{\"u}nsteidl, Berer, Hettich and Veres}]{Gruensteidl2018_v112_p251905}
\bibinfo{author}{Gr{\"u}nsteidl\xfnm[ C.]}, \bibinfo{author}{Berer\xfnm[ T.]}, \bibinfo{author}{Hettich\xfnm[ M.]}, \bibinfo{author}{Veres\xfnm[ I.]}.
\newblock \bibinfo{title}{Determination of thickness and bulk sound velocities of isotropic plates using zero-group-velocity {{Lamb}} waves}.
\newblock \emph{\bibinfo{journal}{Applied Physics Letters}} \bibinfo{year}{2018};\hspace{0pt}\textbf{\bibinfo{volume}{112}}(\bibinfo{number}{25}):\bibinfo{pages}{251905}.
\newblock \DOIprefix\doi{10.1063/1.5034313}.
\bibitem[{Balogun et~al.(2011)Balogun, Cole, Huber, Chinn, Murray and Spicer}]{balogun_high-spatial-resolution_2011}
\bibinfo{author}{Balogun\xfnm[ O.]}, \bibinfo{author}{Cole\xfnm[ G.D.]}, \bibinfo{author}{Huber\xfnm[ R.]}, \bibinfo{author}{Chinn\xfnm[ D.]}, \bibinfo{author}{Murray\xfnm[ T.W.]}, \bibinfo{author}{Spicer\xfnm[ J.B.]}.
\newblock \bibinfo{title}{High-spatial-resolution sub-surface imaging using a laser-based acoustic microscopy technique}.
\newblock \emph{\bibinfo{journal}{IEEE Transactions on Ultrasonics, Ferroelectrics, and Frequency Control}} \bibinfo{year}{2011};\hspace{0pt}\textbf{\bibinfo{volume}{58}}(\bibinfo{number}{1}):\bibinfo{pages}{226--233}.
\newblock \DOIprefix\doi{10.1109/TUFFC.2011.1789}.
\bibitem[{C{\`e}s et~al.(2011{\natexlab{a}})C{\`e}s, Clorennec, Royer and Prada}]{ces_thin_2011}
\bibinfo{author}{C{\`e}s\xfnm[ M.]}, \bibinfo{author}{Clorennec\xfnm[ D.]}, \bibinfo{author}{Royer\xfnm[ D.]}, \bibinfo{author}{Prada\xfnm[ C.]}.
\newblock \bibinfo{title}{Thin layer thickness measurements by zero group velocity {{Lamb}} mode resonances}.
\newblock \emph{\bibinfo{journal}{Review of Scientific Instruments}} \bibinfo{year}{2011}{\natexlab{a}};\hspace{0pt}\textbf{\bibinfo{volume}{82}}(\bibinfo{number}{11}):\bibinfo{pages}{114902}.
\newblock \DOIprefix\doi{10.1063/1.3660182}.
\bibitem[{Prada et~al.(2008{\natexlab{b}})Prada, Clorennec and Royer}]{prada_power_2008}
\bibinfo{author}{Prada\xfnm[ C.]}, \bibinfo{author}{Clorennec\xfnm[ D.]}, \bibinfo{author}{Royer\xfnm[ D.]}.
\newblock \bibinfo{title}{Power law decay of zero group velocity {{Lamb}} modes}.
\newblock \emph{\bibinfo{journal}{Wave Motion}} \bibinfo{year}{2008}{\natexlab{b}};\hspace{0pt}\textbf{\bibinfo{volume}{45}}(\bibinfo{number}{6}):\bibinfo{pages}{723--728}.
\newblock \DOIprefix\doi{10.1016/j.wavemoti.2007.11.005}.
\bibitem[{Yan et~al.(2021)Yan, Raetz, Groby, Duclos, Geslain, Chigarev et~al.}]{yan_estimation_2021}
\bibinfo{author}{Yan\xfnm[ G.]}, \bibinfo{author}{Raetz\xfnm[ S.]}, \bibinfo{author}{Groby\xfnm[ J.P.]}, \bibinfo{author}{Duclos\xfnm[ A.]}, \bibinfo{author}{Geslain\xfnm[ A.]}, \bibinfo{author}{Chigarev\xfnm[ N.]}, et~al.
\newblock \bibinfo{title}{Estimation via {{Laser Ultrasonics}} of the {{Ultrasonic Attenuation}} in a {{Polycrystalline Aluminum Thin Plate Using Complex Wavenumber Recovery}} in the {{Vicinity}} of a {{Zero-Group-Velocity Lamb Mode}}}.
\newblock \emph{\bibinfo{journal}{Applied Sciences}} \bibinfo{year}{2021};\hspace{0pt}\textbf{\bibinfo{volume}{11}}(\bibinfo{number}{15}):\bibinfo{pages}{6924}.
\newblock \DOIprefix\doi{10.3390/app11156924}.
\bibitem[{Ryzy et~al.(2023)Ryzy, Veres, Berer, Salfinger, Kreuzer, Yan et~al.}]{ryzy_determining_2023}
\bibinfo{author}{Ryzy\xfnm[ M.]}, \bibinfo{author}{Veres\xfnm[ I.]}, \bibinfo{author}{Berer\xfnm[ T.]}, \bibinfo{author}{Salfinger\xfnm[ M.]}, \bibinfo{author}{Kreuzer\xfnm[ S.]}, \bibinfo{author}{Yan\xfnm[ G.]}, et~al.
\newblock \bibinfo{title}{Determining longitudinal and transverse elastic wave attenuation from zero-group-velocity {{Lamb}} waves in a pair of plates}.
\newblock \emph{\bibinfo{journal}{The Journal of the Acoustical Society of America}} \bibinfo{year}{2023};\hspace{0pt}\textbf{\bibinfo{volume}{153}}(\bibinfo{number}{4}):\bibinfo{pages}{2090--2100}.
\newblock \DOIprefix\doi{10.1121/10.0017652}.
\bibitem[{Thelen et~al.(2021)Thelen, Bochud, Brinker, Prada and Huber}]{thelen_laser-excited_2021}
\bibinfo{author}{Thelen\xfnm[ M.]}, \bibinfo{author}{Bochud\xfnm[ N.]}, \bibinfo{author}{Brinker\xfnm[ M.]}, \bibinfo{author}{Prada\xfnm[ C.]}, \bibinfo{author}{Huber\xfnm[ P.]}.
\newblock \bibinfo{title}{Laser-excited elastic guided waves reveal the complex mechanics of nanoporous silicon}.
\newblock \emph{\bibinfo{journal}{Nature Communications}} \bibinfo{year}{2021};\hspace{0pt}\textbf{\bibinfo{volume}{12}}(\bibinfo{number}{1}):\bibinfo{pages}{3597}.
\newblock \DOIprefix\doi{10.1038/s41467-021-23398-0}.
\bibitem[{Yan et~al.(2020)Yan, Raetz, Chigarev, Blondeau, Gusev and Tournat}]{Yan2020_v116_p102323}
\bibinfo{author}{Yan\xfnm[ G.]}, \bibinfo{author}{Raetz\xfnm[ S.]}, \bibinfo{author}{Chigarev\xfnm[ N.]}, \bibinfo{author}{Blondeau\xfnm[ J.]}, \bibinfo{author}{Gusev\xfnm[ V.E.]}, \bibinfo{author}{Tournat\xfnm[ V.]}.
\newblock \bibinfo{title}{Cumulative fatigue damage in thin aluminum films evaluated non-destructively with lasers via zero-group-velocity {{Lamb}} modes}.
\newblock \emph{\bibinfo{journal}{NDT \& E International}} \bibinfo{year}{2020};\hspace{0pt}\textbf{\bibinfo{volume}{116}}:\bibinfo{pages}{102323}.
\newblock \DOIprefix\doi{10.1016/j.ndteint.2020.102323}.
\bibitem[{Clorennec et~al.(2007)Clorennec, Prada and Royer}]{clorennec_local_2007}
\bibinfo{author}{Clorennec\xfnm[ D.]}, \bibinfo{author}{Prada\xfnm[ C.]}, \bibinfo{author}{Royer\xfnm[ D.]}.
\newblock \bibinfo{title}{Local and noncontact measurements of bulk acoustic wave velocities in thin isotropic plates and shells using zero group velocity {{Lamb}} modes}.
\newblock \emph{\bibinfo{journal}{Journal of Applied Physics}} \bibinfo{year}{2007};\hspace{0pt}\textbf{\bibinfo{volume}{101}}(\bibinfo{number}{3}):\bibinfo{pages}{034908}.
\newblock \DOIprefix\doi{10.1063/1.2434824}.
\bibitem[{C{\`e}s et~al.(2012)C{\`e}s, Royer and Prada}]{ces_characterization_2012}
\bibinfo{author}{C{\`e}s\xfnm[ M.]}, \bibinfo{author}{Royer\xfnm[ D.]}, \bibinfo{author}{Prada\xfnm[ C.]}.
\newblock \bibinfo{title}{Characterization of mechanical properties of a hollow cylinder with zero group velocity {{Lamb}} modes}.
\newblock \emph{\bibinfo{journal}{The Journal of the Acoustical Society of America}} \bibinfo{year}{2012};\hspace{0pt}\textbf{\bibinfo{volume}{132}}(\bibinfo{number}{1}):\bibinfo{pages}{180--185}.
\newblock \DOIprefix\doi{10.1121/1.4726033}.
\bibitem[{Gr{\"u}nsteidl et~al.(2016)Gr{\"u}nsteidl, Murray, Berer and Veres}]{grunsteidl_inverse_2016}
\bibinfo{author}{Gr{\"u}nsteidl\xfnm[ C.]}, \bibinfo{author}{Murray\xfnm[ T.W.]}, \bibinfo{author}{Berer\xfnm[ T.]}, \bibinfo{author}{Veres\xfnm[ I.A.]}.
\newblock \bibinfo{title}{Inverse characterization of plates using zero group velocity {{Lamb}} modes}.
\newblock \emph{\bibinfo{journal}{Ultrasonics}} \bibinfo{year}{2016};\hspace{0pt}\textbf{\bibinfo{volume}{65}}:\bibinfo{pages}{1--4}.
\newblock \DOIprefix\doi{10.1016/j.ultras.2015.10.015}.
\bibitem[{Watzl et~al.(2022)Watzl, Kerschbaummayr, Schagerl, Mitter, Sonderegger and Gr{\"u}nsteidl}]{watzl_situ_2022}
\bibinfo{author}{Watzl\xfnm[ G.]}, \bibinfo{author}{Kerschbaummayr\xfnm[ C.]}, \bibinfo{author}{Schagerl\xfnm[ M.]}, \bibinfo{author}{Mitter\xfnm[ T.]}, \bibinfo{author}{Sonderegger\xfnm[ B.]}, \bibinfo{author}{Gr{\"u}nsteidl\xfnm[ C.]}.
\newblock \bibinfo{title}{In situ laser-ultrasonic monitoring of {{Poisson}}'s ratio and bulk sound velocities of steel plates during thermal processes}.
\newblock \emph{\bibinfo{journal}{Acta Materialia}} \bibinfo{year}{2022};\hspace{0pt}\textbf{\bibinfo{volume}{235}}:\bibinfo{pages}{118097}.
\newblock \DOIprefix\doi{10.1016/j.actamat.2022.118097}.
\bibitem[{Morales et~al.(2024)Morales, Stobbe, Lum, Kube, Murray and Kube}]{Morales2024}
\bibinfo{author}{Morales\xfnm[ R.E.]}, \bibinfo{author}{Stobbe\xfnm[ D.M.]}, \bibinfo{author}{Lum\xfnm[ J.S.]}, \bibinfo{author}{Kube\xfnm[ C.M.]}, \bibinfo{author}{Murray\xfnm[ T.W.]}, \bibinfo{author}{Kube\xfnm[ C.M.]}.
\newblock \bibinfo{title}{{Acoustoelastic characterization of plates using zero group velocity Lamb modes}}.
\newblock \emph{\bibinfo{journal}{Appl Phys Lett}} \bibinfo{year}{2024};\hspace{0pt}\textbf{\bibinfo{volume}{124}}:\bibinfo{pages}{084101}.
\newblock \DOIprefix\doi{10.1063/5.0183620}.
\bibitem[{Zhang et~al.(2025)Zhang, Hu, Deng and Li}]{Zhang2025}
\bibinfo{author}{Zhang\xfnm[ C.]}, \bibinfo{author}{Hu\xfnm[ Y.]}, \bibinfo{author}{Deng\xfnm[ M.]}, \bibinfo{author}{Li\xfnm[ W.]}.
\newblock \bibinfo{title}{{Zero-group velocity combined-harmonic generation through counter-directional mixing and application for internal damage localization in composites}}.
\newblock \emph{\bibinfo{journal}{Journal of Applied Physics}} \bibinfo{year}{2025};\hspace{0pt}\textbf{\bibinfo{volume}{137}}(\bibinfo{number}{9}).
\newblock \DOIprefix\doi{10.1063/5.0244382}.
\bibitem[{Chen et~al.(2026)Chen, Deng, Chen, Gao and Bai}]{Chen2026}
\bibinfo{author}{Chen\xfnm[ H.]}, \bibinfo{author}{Deng\xfnm[ M.]}, \bibinfo{author}{Chen\xfnm[ Y.]}, \bibinfo{author}{Gao\xfnm[ G.]}, \bibinfo{author}{Bai\xfnm[ Y.]}.
\newblock \bibinfo{title}{{Modeling and simulation of zero-group-velocity combined harmonic generated by two counter-directional Lamb waves mixing in an adhesively bonded plate}}.
\newblock \emph{\bibinfo{journal}{Ultrasonics}} \bibinfo{year}{2026};\hspace{0pt}\textbf{\bibinfo{volume}{160}}(\bibinfo{number}{November 2025}):\bibinfo{pages}{107889}.
\newblock \DOIprefix\doi{10.1016/j.ultras.2025.107889}.
\bibitem[{Lu and Shen(2025)}]{Lu2025}
\bibinfo{author}{Lu\xfnm[ R.]}, \bibinfo{author}{Shen\xfnm[ Y.]}.
\newblock \bibinfo{title}{{Zero group velocity mode nonlinear ultrasonics for fatigue crack detection}}.
\newblock \emph{\bibinfo{journal}{Ultrasonics}} \bibinfo{year}{2025};\hspace{0pt}\textbf{\bibinfo{volume}{150}}(\bibinfo{number}{January}):\bibinfo{pages}{107604}.
\newblock \DOIprefix\doi{10.1016/j.ultras.2025.107604}.
\bibitem[{Lamb(1882)}]{lamb_vibrations_1882}
\bibinfo{author}{Lamb\xfnm[ H.]}.
\newblock \bibinfo{title}{On the vibrations of a spherical shell}.
\newblock \emph{\bibinfo{journal}{Proceedings of the London Mathematical Society}} \bibinfo{year}{1882};\hspace{0pt}\textbf{\bibinfo{volume}{s1-14}}(\bibinfo{number}{1}):\bibinfo{pages}{50--56}.
\newblock \DOIprefix\doi{10.1112/plms/s1-14.1.50}.
\bibitem[{Towfighi and Kundu(2003)}]{towfighi_elastic_2003}
\bibinfo{author}{Towfighi\xfnm[ S.]}, \bibinfo{author}{Kundu\xfnm[ T.]}.
\newblock \bibinfo{title}{Elastic wave propagation in anisotropic spherical curved plates}.
\newblock \emph{\bibinfo{journal}{International Journal of Solids and Structures}} \bibinfo{year}{2003};\hspace{0pt}\textbf{\bibinfo{volume}{40}}(\bibinfo{number}{20}):\bibinfo{pages}{5495--5510}.
\newblock \DOIprefix\doi{10.1016/S0020-7683(03)00278-6}.
\bibitem[{Scruby and Drain(1990{\natexlab{b}})}]{scruby_5_1990}
\bibinfo{author}{Scruby\xfnm[ C.B.]}, \bibinfo{author}{Drain\xfnm[ L.E.]}.
\newblock \bibinfo{title}{5 {{Ultrasonic Generation}} by {{Laser}}}.
\newblock In: \emph{\bibinfo{booktitle}{Laser {{Ultrasonics Techniques}} and {{Applications}}}}. \bibinfo{publisher}{Adam Hilger}.
\newblock ISBN \bibinfo{isbn}{0-7503-0050-7}; \bibinfo{year}{1990}{\natexlab{b}}, \hspace{0pt}p. \bibinfo{pages}{223--324}.
\bibitem[{Grahn et~al.(1989)Grahn, Maris and Tauc}]{grahn_picosecond_1989}
\bibinfo{author}{Grahn\xfnm[ H.]}, \bibinfo{author}{Maris\xfnm[ H.]}, \bibinfo{author}{Tauc\xfnm[ J.]}.
\newblock \bibinfo{title}{Picosecond ultrasonics}.
\newblock \emph{\bibinfo{journal}{IEEE Journal of Quantum Electronics}} \bibinfo{year}{1989};\hspace{0pt}\textbf{\bibinfo{volume}{25}}(\bibinfo{number}{12}):\bibinfo{pages}{2562--2569}.
\newblock \DOIprefix\doi{10.1109/3.40643}.
\bibitem[{Matsuda et~al.(2015)Matsuda, Larciprete, Li~Voti and Wright}]{matsuda_fundamentals_2015}
\bibinfo{author}{Matsuda\xfnm[ O.]}, \bibinfo{author}{Larciprete\xfnm[ M.C.]}, \bibinfo{author}{Li~Voti\xfnm[ R.]}, \bibinfo{author}{Wright\xfnm[ O.B.]}.
\newblock \bibinfo{title}{Fundamentals of picosecond laser ultrasonics}.
\newblock \emph{\bibinfo{journal}{Ultrasonics}} \bibinfo{year}{2015};\hspace{0pt}\textbf{\bibinfo{volume}{56}}:\bibinfo{pages}{3--20}.
\newblock \DOIprefix\doi{10.1016/j.ultras.2014.06.005}.
\bibitem[{Murray and Balogun(2004)}]{murray_high-sensitivity_2004}
\bibinfo{author}{Murray\xfnm[ T.W.]}, \bibinfo{author}{Balogun\xfnm[ O.]}.
\newblock \bibinfo{title}{High-sensitivity laser-based acoustic microscopy using a modulated excitation source}.
\newblock \emph{\bibinfo{journal}{Applied Physics Letters}} \bibinfo{year}{2004};\hspace{0pt}\textbf{\bibinfo{volume}{85}}(\bibinfo{number}{14}):\bibinfo{pages}{2974--2976}.
\newblock \DOIprefix\doi{10.1063/1.1802387}.
\bibitem[{Ryzy et~al.(2018{\natexlab{a}})Ryzy, Grabec, {\"O}sterreicher, Hettich and Veres}]{ryzy_measurement_2018}
\bibinfo{author}{Ryzy\xfnm[ M.]}, \bibinfo{author}{Grabec\xfnm[ T.]}, \bibinfo{author}{{\"O}sterreicher\xfnm[ J.A.]}, \bibinfo{author}{Hettich\xfnm[ M.]}, \bibinfo{author}{Veres\xfnm[ I.A.]}.
\newblock \bibinfo{title}{Measurement of coherent surface acoustic wave attenuation in polycrystalline aluminum}.
\newblock \emph{\bibinfo{journal}{AIP Advances}} \bibinfo{year}{2018}{\natexlab{a}};\hspace{0pt}\textbf{\bibinfo{volume}{8}}(\bibinfo{number}{12}):\bibinfo{pages}{125019}.
\newblock \DOIprefix\doi{10.1063/1.5074180}.
\bibitem[{Gr{\"u}nsteidl et~al.(2020{\natexlab{a}})Gr{\"u}nsteidl, Ryzy, Hettich, Berer and Veres}]{grunsteidl_evaluation_2020-1}
\bibinfo{author}{Gr{\"u}nsteidl\xfnm[ C.]}, \bibinfo{author}{Ryzy\xfnm[ M.]}, \bibinfo{author}{Hettich\xfnm[ M.]}, \bibinfo{author}{Berer\xfnm[ T.]}, \bibinfo{author}{Veres\xfnm[ I.]}.
\newblock \bibinfo{title}{Evaluation of elastic wave attenuation in the {{GHz}} range using zero-group-velocity resonances}.
\newblock In: \emph{\bibinfo{booktitle}{Proceedings of the {{Forum Acusticum}}}}. \bibinfo{address}{Lyon, France}: \bibinfo{publisher}{e-Forum Acusticum 2020}; \bibinfo{year}{2020}{\natexlab{a}}, \hspace{0pt}p. \bibinfo{pages}{2557--2560}.
\newblock \DOIprefix\doi{10.48465/FA.2020.0582}.
\bibitem[{Gr{\"u}nsteidl et~al.(2020{\natexlab{b}})Gr{\"u}nsteidl, Veres, Berer, Kreuzer, Rothemund, Hettich et~al.}]{grunsteidl_measurement_2020}
\bibinfo{author}{Gr{\"u}nsteidl\xfnm[ C.]}, \bibinfo{author}{Veres\xfnm[ I.]}, \bibinfo{author}{Berer\xfnm[ T.]}, \bibinfo{author}{Kreuzer\xfnm[ S.]}, \bibinfo{author}{Rothemund\xfnm[ R.]}, \bibinfo{author}{Hettich\xfnm[ M.]}, et~al.
\newblock \bibinfo{title}{Measurement of the attenuation of elastic waves at {{GHz}} frequencies using resonant thickness modes}.
\newblock \emph{\bibinfo{journal}{Applied Physics Letters}} \bibinfo{year}{2020}{\natexlab{b}};\hspace{0pt}\textbf{\bibinfo{volume}{117}}(\bibinfo{number}{16}):\bibinfo{pages}{164102}.
\newblock \DOIprefix\doi{10.1063/5.0026367}.
\bibitem[{noa(2025)}]{noauthor_diamond_2025}
\bibinfo{title}{Diamond {{Materials GmbH}}, {{Feiburg}}, {{Germany}}}.
\newblock \bibinfo{howpublished}{https://www.diamond-materials.com/en/cvd-diamond/mechanical/}; \bibinfo{year}{2025}.
\bibitem[{Klein and Cardinale(1993)}]{Klein1993diamond}
\bibinfo{author}{Klein\xfnm[ C.A.]}, \bibinfo{author}{Cardinale\xfnm[ G.F.]}.
\newblock \bibinfo{title}{Young's modulus and poisson's ratio of cvd diamond}.
\newblock \emph{\bibinfo{journal}{Diamond and Related Materials}} \bibinfo{year}{1993};\hspace{0pt}\textbf{\bibinfo{volume}{2}}(\bibinfo{number}{5}):\bibinfo{pages}{918--923}.
\newblock \DOIprefix\doi{10.1016/0925-9635(93)90250-6}.
\bibitem[{Cho et~al.(2005)Cho, Chasiotis, Friedmann and Sullivan}]{Cho2005diamond}
\bibinfo{author}{Cho\xfnm[ S.]}, \bibinfo{author}{Chasiotis\xfnm[ I.]}, \bibinfo{author}{Friedmann\xfnm[ T.A.]}, \bibinfo{author}{Sullivan\xfnm[ J.P.]}.
\newblock \bibinfo{title}{Young's modulus, poisson's ratio and failure properties of tetrahedral amorphous diamond-like carbon for mems devices}.
\newblock \emph{\bibinfo{journal}{Journal of Micromechanics and Microengineering}} \bibinfo{year}{2005};\hspace{0pt}\textbf{\bibinfo{volume}{15}}(\bibinfo{number}{4}):\bibinfo{pages}{728}.
\newblock \DOIprefix\doi{10.1088/0960-1317/15/4/009}.
\bibitem[{Relyea et~al.(1998)Relyea, White, McGrath and Beck}]{Relyea1998}
\bibinfo{author}{Relyea\xfnm[ H.]}, \bibinfo{author}{White\xfnm[ M.]}, \bibinfo{author}{McGrath\xfnm[ J.]}, \bibinfo{author}{Beck\xfnm[ J.]}.
\newblock \bibinfo{title}{Thermal diffusivity measurements of free-standing cvd diamond films using non-contacting, non-destructive techniques}.
\newblock \emph{\bibinfo{journal}{Diamond and Related Materials}} \bibinfo{year}{1998};\hspace{0pt}\textbf{\bibinfo{volume}{7}}(\bibinfo{number}{8}):\bibinfo{pages}{1207--1212}.
\newblock \DOIprefix\doi{10.1016/S0925-9635(98)00183-6}.
\bibitem[{Scruby and Drain(1990{\natexlab{c}})}]{scrubyLaserUltrasonicsTechniques1990_p76}
\bibinfo{author}{Scruby\xfnm[ C.B.]}, \bibinfo{author}{Drain\xfnm[ L.E.]}.
\newblock \bibinfo{title}{3.2.1 {{The Michelson}} interferometer}.
\newblock In: \emph{\bibinfo{booktitle}{Laser {{Ultrasonics Techniques}} and {{Applications}}}}. \bibinfo{publisher}{Adam Hilger}.
\newblock ISBN \bibinfo{isbn}{0-7503-0050-7}; \bibinfo{year}{1990}{\natexlab{c}}, \hspace{0pt}p.~\bibinfo{pages}{76}.
\bibitem[{Kiefer(2023)}]{kiefer_gewtool_2023}
\bibinfo{author}{Kiefer\xfnm[ D.]}.
\newblock \bibinfo{title}{{{GEWtool}} [{{Computer}} software]}.
\newblock \bibinfo{year}{2023}.
\newblock \DOIprefix\doi{10.5281/zenodo.10114243}.
\bibitem[{Kiefer et~al.(2023)Kiefer, Plestenjak, Gravenkamp and Prada}]{kiefer_computing_2023}
\bibinfo{author}{Kiefer\xfnm[ D.A.]}, \bibinfo{author}{Plestenjak\xfnm[ B.]}, \bibinfo{author}{Gravenkamp\xfnm[ H.]}, \bibinfo{author}{Prada\xfnm[ C.]}.
\newblock \bibinfo{title}{Computing zero-group-velocity points in anisotropic elastic waveguides: {{Globally}} and locally convergent methods}.
\newblock \emph{\bibinfo{journal}{The Journal of the Acoustical Society of America}} \bibinfo{year}{2023};\hspace{0pt}\textbf{\bibinfo{volume}{153}}(\bibinfo{number}{2}):\bibinfo{pages}{1386--1398}.
\newblock \DOIprefix\doi{10.1121/10.0017252}.
\bibitem[{Kiefer et~al.(2025)Kiefer, Watzl, Burgholzer, Ryzy and Gr{\"u}nsteidl}]{kiefer_electroelastic_2025}
\bibinfo{author}{Kiefer\xfnm[ D.A.]}, \bibinfo{author}{Watzl\xfnm[ G.]}, \bibinfo{author}{Burgholzer\xfnm[ K.]}, \bibinfo{author}{Ryzy\xfnm[ M.]}, \bibinfo{author}{Gr{\"u}nsteidl\xfnm[ C.]}.
\newblock \bibinfo{title}{Electroelastic guided wave dispersion in piezoelectric plates: {{Spectral}} methods and laser-ultrasound experiments}.
\newblock \emph{\bibinfo{journal}{Journal of Applied Physics}} \bibinfo{year}{2025};\hspace{0pt}\textbf{\bibinfo{volume}{137}}(\bibinfo{number}{11}):\bibinfo{pages}{114502}.
\newblock \DOIprefix\doi{10.1063/5.0250494}.
\bibitem[{Every et~al.(2013)Every, Utegulov and Veres}]{every_laser_2013}
\bibinfo{author}{Every\xfnm[ A.G.]}, \bibinfo{author}{Utegulov\xfnm[ Z.N.]}, \bibinfo{author}{Veres\xfnm[ I.A.]}.
\newblock \bibinfo{title}{Laser thermoelastic generation in metals above the melt threshold}.
\newblock \emph{\bibinfo{journal}{Journal of Applied Physics}} \bibinfo{year}{2013};\hspace{0pt}\textbf{\bibinfo{volume}{114}}(\bibinfo{number}{20}):\bibinfo{pages}{203508}.
\newblock \DOIprefix\doi{10.1063/1.4832483}.
\bibitem[{Ryzy et~al.(2018{\natexlab{b}})Ryzy, Grabec, Sedl{\'a}k and Veres}]{Ryzy2018_v143_p219}
\bibinfo{author}{Ryzy\xfnm[ M.]}, \bibinfo{author}{Grabec\xfnm[ T.]}, \bibinfo{author}{Sedl{\'a}k\xfnm[ P.]}, \bibinfo{author}{Veres\xfnm[ I.A.]}.
\newblock \bibinfo{title}{Influence of grain morphology on ultrasonic wave attenuation in polycrystalline media with statistically equiaxed grains}.
\newblock \emph{\bibinfo{journal}{The Journal of the Acoustical Society of America}} \bibinfo{year}{2018}{\natexlab{b}};\hspace{0pt}\textbf{\bibinfo{volume}{143}}(\bibinfo{number}{1}):\bibinfo{pages}{219--229}.
\newblock \DOIprefix\doi{10.1121/1.5020785}.
\bibitem[{Grabec et~al.(2022)Grabec, Veres and Ryzy}]{grabec_surface_2022}
\bibinfo{author}{Grabec\xfnm[ T.]}, \bibinfo{author}{Veres\xfnm[ I.A.]}, \bibinfo{author}{Ryzy\xfnm[ M.]}.
\newblock \bibinfo{title}{Surface acoustic wave attenuation in polycrystals: {{Numerical}} modeling using a statistical digital twin of an actual sample}.
\newblock \emph{\bibinfo{journal}{Ultrasonics}} \bibinfo{year}{2022};\hspace{0pt}\textbf{\bibinfo{volume}{119}}:\bibinfo{pages}{106585}.
\newblock \DOIprefix\doi{10.1016/j.ultras.2021.106585}.
\bibitem[{{Abu-Shawareb} et~al.(2022){Abu-Shawareb}, Acree, Adams, Adams, Addis, Aden et~al.}]{abu-shawarebLawsonCriterionIgnition2022}
\bibinfo{author}{{Abu-Shawareb}\xfnm[ H.]}, \bibinfo{author}{Acree\xfnm[ R.]}, \bibinfo{author}{Adams\xfnm[ P.]}, \bibinfo{author}{Adams\xfnm[ J.]}, \bibinfo{author}{Addis\xfnm[ B.]}, \bibinfo{author}{Aden\xfnm[ R.]}, et~al.
\newblock \bibinfo{title}{Lawson {{Criterion}} for {{Ignition Exceeded}} in an {{Inertial Fusion Experiment}}}.
\newblock \emph{\bibinfo{journal}{Physical Review Letters}} \bibinfo{year}{2022};\hspace{0pt}\textbf{\bibinfo{volume}{129}}(\bibinfo{number}{7}):\bibinfo{pages}{075001}.
\newblock \DOIprefix\doi{10.1103/PhysRevLett.129.075001}.
\bibitem[{Shui et~al.(1988)Shui, Royer, Dieulesaint and Sun}]{shui_resonance_1988}
\bibinfo{author}{Shui\xfnm[ Y.]}, \bibinfo{author}{Royer\xfnm[ D.]}, \bibinfo{author}{Dieulesaint\xfnm[ E.]}, \bibinfo{author}{Sun\xfnm[ Z.]}.
\newblock \bibinfo{title}{Resonance of surface waves on spheres}.
\newblock In: \emph{\bibinfo{booktitle}{{{IEEE}} 1988 {{Ultrasonics Symposium Proceedings}}.}} \bibinfo{address}{Chicago, IL, USA}: \bibinfo{publisher}{IEEE}; \bibinfo{year}{1988}, \hspace{0pt}p. \bibinfo{pages}{343--346}.
\newblock \DOIprefix\doi{10.1109/ULTSYM.1988.49396}.
\bibitem[{Royer and Clorennec(2008)}]{royer_theoretical_2008}
\bibinfo{author}{Royer\xfnm[ D.]}, \bibinfo{author}{Clorennec\xfnm[ D.]}.
\newblock \bibinfo{title}{Theoretical and {{Experimental Investigation}} of {{Rayleigh Waves}} on {{Spherical}} and {{Cylindrical Surfaces}}}.
\newblock In: \emph{\bibinfo{booktitle}{1st {{International Symposium}} on {{Laser Ultrasonics}}: {{Science}}, {{Technology}} and {{Applications}}}}. \bibinfo{address}{Montreal, Canada}; \bibinfo{year}{2008}, \hspace{0pt}p.~\bibinfo{pages}{8}.
\bibitem[{Yan et~al.(2025)Yan, Huang, Lapa, Squoia, Gr{\"u}nsteidl and Ryzy}]{yan_investigation_2025}
\bibinfo{author}{Yan\xfnm[ G.]}, \bibinfo{author}{Huang\xfnm[ H.]}, \bibinfo{author}{Lapa\xfnm[ P.]}, \bibinfo{author}{Squoia\xfnm[ K.]}, \bibinfo{author}{Gr{\"u}nsteidl\xfnm[ C.]}, \bibinfo{author}{Ryzy\xfnm[ M.]}.
\newblock \bibinfo{title}{Investigation of {{Zero-Group-Velocity Resonance}} in {{Inertial Confinement Fusion Target Capsules}} with {{Frequency-domain Laser Ultrasound Microscopy}}}.
\newblock In: \emph{\bibinfo{booktitle}{Journal of {{Physics}}: {{Conference Series}}}}; vol. \bibinfo{volume}{2966}. \bibinfo{year}{2025}, \hspace{0pt}p. \bibinfo{pages}{012007}.
\newblock \DOIprefix\doi{10.1088/1742-6596/2966/1/012007}.
\bibitem[{Mace and Manconi(2012)}]{Mace2012}
\bibinfo{author}{Mace\xfnm[ B.R.]}, \bibinfo{author}{Manconi\xfnm[ E.]}.
\newblock \bibinfo{title}{Wave motion and dispersion phenomena: Veering, locking and strong coupling effects}.
\newblock \emph{\bibinfo{journal}{The Journal of the Acoustical Society of America}} \bibinfo{year}{2012};\hspace{0pt}\textbf{\bibinfo{volume}{131}}:\bibinfo{pages}{1015--1028}.
\newblock \DOIprefix\doi{10.1121/1.3672647}.
\bibitem[{Veres et~al.(2014)Veres, Berer, Gr\"{u}nsteidl and Burgholzer}]{Veres2014crossingPoints}
\bibinfo{author}{Veres\xfnm[ I.A.]}, \bibinfo{author}{Berer\xfnm[ T.]}, \bibinfo{author}{Gr\"{u}nsteidl\xfnm[ C.]}, \bibinfo{author}{Burgholzer\xfnm[ P.]}.
\newblock \bibinfo{title}{On the crossing points of the lamb modes and the maxima and minima of displacements observed at the surface}.
\newblock \emph{\bibinfo{journal}{Ultrasonics}} \bibinfo{year}{2014};\hspace{0pt}\textbf{\bibinfo{volume}{54}}(\bibinfo{number}{3}):\bibinfo{pages}{759--762}.
\newblock \DOIprefix\doi{10.1016/j.ultras.2013.10.018}.
\bibitem[{Gravenkamp et~al.(2023)Gravenkamp, Plestenjak and Kiefer}]{Gravenkamp2023}
\bibinfo{author}{Gravenkamp\xfnm[ H.]}, \bibinfo{author}{Plestenjak\xfnm[ B.]}, \bibinfo{author}{Kiefer\xfnm[ D.A.]}.
\newblock \bibinfo{title}{Notes on osculations and mode tracing in semi-analytical waveguide modeling}.
\newblock \emph{\bibinfo{journal}{Ultrasonics}} \bibinfo{year}{2023};\hspace{0pt}\textbf{\bibinfo{volume}{135}}:\bibinfo{pages}{107112}.
\newblock \DOIprefix\doi{10.1016/j.ultras.2023.107112}.
\bibitem[{Rose(2004)}]{rose_ultrasonic_2004}
\bibinfo{author}{Rose\xfnm[ J.L.]}.
\newblock \emph{\bibinfo{title}{Ultrasonic {{Waves}} in {{Solid Media}}}}.
\newblock \bibinfo{publisher}{Cambridge University Press}; \bibinfo{year}{2004}.
\newblock ISBN \bibinfo{isbn}{0 521 54889 6}.
\bibitem[{C{\`e}s et~al.(2011{\natexlab{b}})C{\`e}s, Clorennec, Royer and Prada}]{ces_edge_2011}
\bibinfo{author}{C{\`e}s\xfnm[ M.]}, \bibinfo{author}{Clorennec\xfnm[ D.]}, \bibinfo{author}{Royer\xfnm[ D.]}, \bibinfo{author}{Prada\xfnm[ C.]}.
\newblock \bibinfo{title}{Edge resonance and zero group velocity {{Lamb}} modes in a free elastic plate}.
\newblock \emph{\bibinfo{journal}{The Journal of the Acoustical Society of America}} \bibinfo{year}{2011}{\natexlab{b}};\hspace{0pt}\textbf{\bibinfo{volume}{130}}(\bibinfo{number}{2}):\bibinfo{pages}{689--694}.
\newblock \DOIprefix\doi{10.1121/1.3607417}.

\end{thebibliography}

\end{document}